\documentclass[11pt,a4paper]{article}
\usepackage[x11names, rgb, html]{xcolor}

\definecolor{red}{HTML}{f54b1a}
\definecolor{pink}{HTML}{d19eb1}
\definecolor{orange}{HTML}{d3772e}
\definecolor{yellow}{HTML}{ebe85d}
\definecolor{green}{HTML}{0f6852}
\definecolor{lightblue}{HTML}{01abe9}
\definecolor{darkblue}{HTML}{1b346c}
\definecolor{tan}{HTML}{e5c39e}
\definecolor{darktan}{HTML}{af9e73}
\definecolor{grey}{HTML}{c3ced0}
\definecolor{darkgrey}{HTML}{9dadc4}
\definecolor{black}{HTML}{110d1b}
\definecolor{white}{HTML}{f1f8f1}

\usepackage{hyperref}
\hypersetup{
  colorlinks = true,
  linkcolor = red,
  citecolor = darkblue,
  urlcolor = lightblue
}

\usepackage[cmintegrals,cmbraces]{newtxmath}

\usepackage{graphicx}

\usepackage[margin=1.25in]{geometry}

\usepackage{algorithm}
\usepackage[noend]{algpseudocode}
\algrenewcommand{\algorithmiccomment}[1]{$\vartriangleright$ #1}
\algrenewcommand{\algorithmicreturn}{\textbf{Return: }}
\algnewcommand\algorithmicinput{\textbf{Input: }}
\algnewcommand\Input{\State \algorithmicinput}

\newtheorem{theorem}{Theorem}[section]
\newtheorem{proposition}[theorem]{Proposition}
\newtheorem{definition}[theorem]{Definition}

\def\nb{\boldsymbol{n}}

\def\kb{k_{\rm B}}

\def\mb{\boldsymbol{m}}
\def\xb{\boldsymbol{x}}

\def\zb{\boldsymbol{z}}

\def\Mb{\boldsymbol{M}}

\def\grad{\nabla}

\def\RR{\mathbb{R}}

\newcommand{\avg}[1]{\left\langle #1 \right\rangle}

\def\<{\langle} \def\>{\rangle}

\begin{document}


\author{
Shriram Chennakesavalu, David J. Toomer, Grant M. Rotskoff \\ 
\texttt{\{shriramc, djtoomer, rotskoff\}@stanford.edu} \\
Department of Chemistry, Stanford University, Stanford, CA 94305
}
\title{Ensuring thermodynamic consistency with invertible coarse-graining}
\date{\today}

\maketitle

\abstract{Coarse-grained models are a core computational tool in theoretical chemistry and biophysics. A judicious choice of a coarse-grained model can yield physical insight by isolating the essential degrees of freedom that dictate the thermodynamic properties of a complex, condensed-phase system. The reduced complexity of the model typically leads to lower computational costs and more efficient sampling compared to atomistic models. 
Designing ``good'' coarse-grained models is an art. Generally, the mapping from fine-grained configurations to coarse-grained configurations itself is not optimized in any way; instead, the energy function associated with the mapped configurations is. In this work, we explore the consequences of optimizing the coarse-grained representation alongside its potential energy function. We use a graph machine learning framework to embed atomic configurations into a low dimensional space to produce efficient representations of the original molecular system. Because the representation we obtain is no longer directly interpretable as a real space representation of the atomic coordinates, we also introduce an inversion process and an associated thermodynamic consistency relation that allows us to rigorously sample fine-grained configurations conditioned on the coarse-grained sampling. We show that this technique is robust, recovering the first two moments of the distribution of several observables in proteins such as chignolin and alanine dipeptide.}


\section*{Introduction}

Biophysical systems evolve with an intricately orchestrated dynamics, and even the most subtle molecular motions can inform both their large-scale static and dynamic properties.
In most biomolecular systems of interest, there is no reliable way to determine which degrees of freedom can be neglected to obtain an effective model that makes predictions in quantitative agreement with atomistic models.
The coupling of both time and spatial scales creates inherent challenges for molecular simulation: many phenomena we would like to simulate, such as protein conformational change~\cite{pak_advances_2018, noid_multiscale_2008}, protein folding~\cite{clementi_coarsegrained_2008}, and multicomponent self-assembly~\cite{saunders_coarsegraining_2012} occur rarely, requiring simulations far too costly for even the most powerful computers. 
The need to access conformational dynamics on very long timescales has spurred the development of many accelerated sampling methods, which can ameliorate this issue. 
However, these methods typically require defining a low-dimensional coordinate of interest and sampling that low-dimensional space exhaustively. 

Coarse-graining, also known as dimensionality reduction, is intended to provide a model of reduced complexity that can be used, in principle, to accelerate sampling~\cite{izvekov_multiscale_2005, izvekov_multiscale_2005a, noid_multiscale_2008,wang_coarsegraining_2019,wang_machine_2019, husic_coarse_2020}.
For biophysical systems, coarse-grained models are typically developed intuitively by assigning groups of atoms within a molecule to a fixed ``bead'' that represents a salient substructure~\cite{izvekov_multiscale_2005}.
With this representation, the coarse-grained model can be parameterized with a potential energy function of essentially the same functional form as that of the original molecular model.
This strategy has been enormously successful in a variety of contexts~\cite{clementi_coarsegrained_2008,pak_advances_2018,das_lowdimensional_2006}.
However, due to the dimensionality reduction, there are questions that simply cannot be answered using a coarse-grained model, no matter how accurately it has been parameterized.

In this work, we ask if it is possible to accelerate dynamics through dimensionality reduction while maintaining the ability to evaluate equilibrium averages of observables defined on the fine-grained system with quantitative accuracy.
While other works have sought to invert coarse-grained representation, generally they do so in a data-driven way that does not rigorously yield a physical distribution of states~\cite{walther_multimodal_2020, wang_coarsegraining_2019}.
To carry out this procedure, we combine an embedding strategy based on hierarchical dimensionality reduction for graph data~\cite{Ying_DiffPool_2018} with a back-mapping procedure that allows us to rigorously sample Boltzmann weighted configurations of the fine-grained system conditioned on the coarse-grained configurations.
This strategy ensures that we can evaluate averages of arbitrary atomistic observables, some of which cannot even be defined for the corresponding coarse-grained model. 

To construct the coarse-graining map, we do not just specify \emph{a priori} how to embed the atomic coordinates in a low-dimensional space, we instead optimize this mapping.
This change requires a distinct paradigm for coarse-graining, in which we use a state-dependent embedding map that allows for a more flexible, but nonlinear representation of the coarse-grained space.
We simultaneously train the coarse-graining map, its associated potential energy function in the coarse-grained space, and a map that inverts the low-dimensional configurations and conditionally generates new fine-grained structures.

Using generative machine learning models to find low dimensional representations has shown success in biomolecular systems~\cite{wang_gen_cg_2021, kohler_smooth_nfs_2021, noe_boltzmann_2019, wang_data_2022, kohler_flowmatching_2022, mahmoud_accurate_2022}, but inverting a coarse-graining map requires specific structure to ensure that sampling will asymptotically converge to the fine-grained Boltzmann distribution.
We represent the inversion map using normalizing flows~\cite{tabak_density_2010,rezende_variational_2015,papamakarios_normalizing_2021}, which have shown promise for augmenting Markov chain Monte Carlo (MCMC) sampling~\cite{albergo_flowbased_2019,gabrie_adaptive_2022,gabrie_efficient_2021,noe_boltzmann_2019}.
While normalizing flows are challenging to optimize for sampling high-dimensional distributions with multiple metastable states~\cite{kohler_smooth_nfs_2021, gabrie_adaptive_2022}, the conditioned sampling procedure that we employ is considerably easier to train. 

This work weaves together many threads being pursued independently in the machine learning literature and in molecular simulation.
However, we see the present work as not the introduction of an algorithm or computational procedure, but rather a conceptual development in coarse-grained modeling.
We believe that allowing for a more complicated, less interpretable coarse-grained space provides new opportunities to accelerate sampling in the fine-grained space.
The extended notion of weak thermodynamic consistency that we introduce provides a framework on which to build new coarse-graining strategies that are targeted to particular classes of observables, which may allow for more efficient models for precise scientific questions.

\section{Weak formulation of thermodynamic consistency}

Atomic resolution molecular models of biophysical systems can provide detailed and accurate insight into the static and dynamic properties of biomolecules, provided that there are sufficiently powerful computational resources to collect a statistically representative sample of configurations of an $n$-particle system, $\{ \xb_i \}_{i=1}^N.$
Due to ergodicity, an MD simulation in the canonical ensemble samples a Boltzmann distribution, and the probability of a given configuration is given by the familiar expression
\begin{equation}
    p(\xb)d\xb = Z^{-1} e^{-\beta U(\xb)} d\xb,
\end{equation}
where $Z$ is the partition function, $\beta = \tfrac{1}{\kb T}$ is the reduced inverse temperature, and $U:\RR^{3n}~\to~\RR$ is the fine-grained potential energy function.

When we coarse-grain a molecular system, we reduce the dimensionality and necessarily destroy information. 
Hence, when carrying out this destructive process, we should ideally preserve the most important degrees of freedom required to describe the fluctuations of the system. 
In landmark work, Noid et al.~\cite{noid_multiscale_2008} established the notion of \textit{thermodynamic consistency} to provide a formal description of the requirements of a ``good'' coarse-graining map.
In their formulation, we require equivalence between the potential of mean force $\hat F$ and the effective coarse-grained potential $\hat U$,
\begin{equation}
        \hat F(\zb) \equiv -\beta^{-1} \log Z^{-1} \int_{\Omega} e^{-\beta U(\xb)} \delta(\Theta(\xb) - \zb) d\xb \leftrightarrow \hat U(\zb).
\end{equation}
This ensures that canonically distributed samples of coarse-grained configurations, sampled in proportion to 
\begin{equation}
    \hat{Z}^{-1} e^{-\beta \hat U(z)} \equiv \hat \rho(\zb) 
\end{equation}
will recover the projected distribution.
The key observation is that equivalence is defined in the coarse-grained space. 

\begin{figure}[t]
    \centering
    \includegraphics[width=1.0\linewidth]{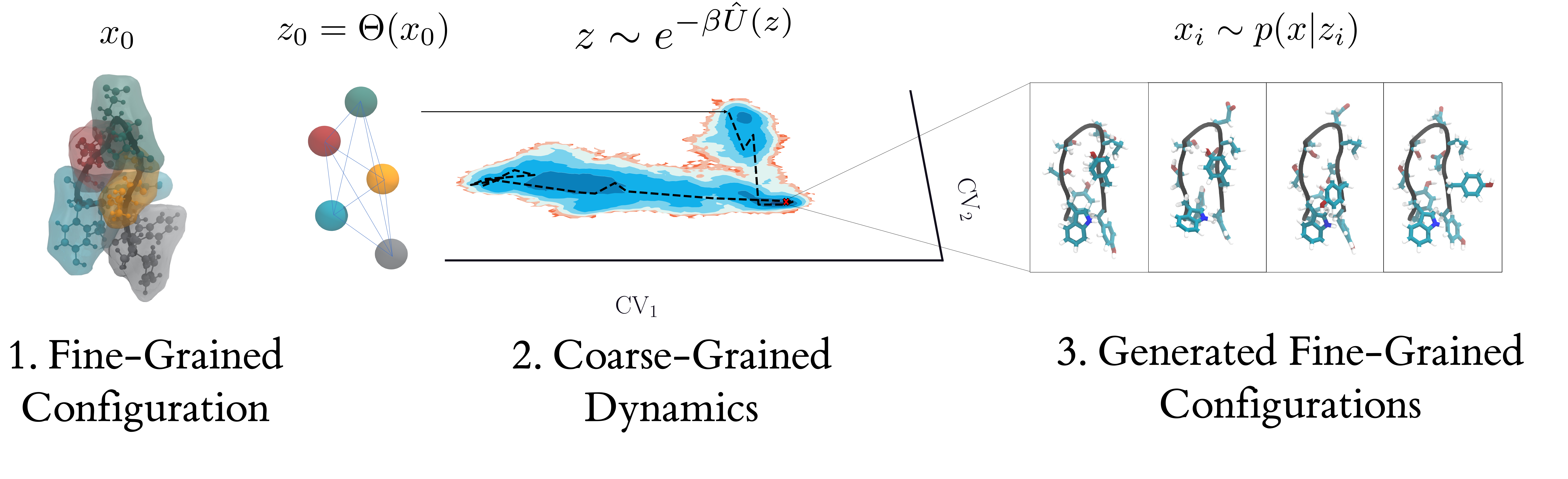}
    \caption{A schematic overview of the coarse-graining procedure. First, a fine-grained molecular structure is embedded with a state-dependent learned projection $\Theta$. A coarse-grained potential $\hat U$ is used to sample configurations coarse-grained configurations $\zb$ so that they are distributed according to a Boltzmann distribution with respect to $\hat U$, as described in Sec.~\ref{sec:embed}. These samples are subsequently used to conditionally sample fine-grained configurations in Sec.~\ref{sec:invert}}
    \label{fig:cg_schematic}
\end{figure}

The requirement of thermodynamic consistency is a stringent one in the sense that it enforces equivalence at the level of the distribution rather than, for example, equivalence of some collection of observables. 
Of course, there are other potentially useful notions of equivalence for probability distributions that could lead to additional flexibility in the procedure.
Here, we take inspiration from the measure theoretic notion of weak convergence, which quantifies the difference between distributions through expectations (or average values) of bounded, continuous functions.
Importantly, this notion could be applied in either the coarse-grained space or the fine-grained space to optimize and test a coarse-graining map.
Throughout, we measure ``weak thermodynamic consistency'' in the fine-grained space.  
We refer to a coarse-graining map $\Theta$ and the associated potential $\hat U$ as ``$\mathcal{F}$ thermodynamically consistent'' if for every observable $f\in \mathcal{F}$,
\begin{equation}
    \hat{Z}^{-1} \int f(\xb) p_{\rm gen}(\xb| \zb) \hat \rho(\zb)\ d\xb d\zb \longrightarrow  Z^{-1} \int f(\xb) \rho(\xb)\ d\xb.
\end{equation}
In this expression, $p_{\rm gen}(\xb | \zb)$ is the conditional probability of generating $\xb$ from a coarse-grained configuration $\zb$---finding a map that performs this inversion in a way that is suitable to reweighting or Monte Carlo is a central goal of the present work and is discussed at length in Sec.~\ref{sec:invert}.
This definition of thermodynamic consistency differs from Ref.~\cite{noid_multiscale_2008} because we only require equivalence on some set of observables $\mathcal{F}$, which could be adapted to a particular problem.

In the appendix, we prove the following straightforward proposition, which relates weak thermodynamic consistency to the definition introduced by Voth and Noid~\cite{noid_multiscale_2008}. 
\begin{proposition}
Let $(\Theta, \hat U, T)$ be an invertible coarse-graining. 
Let $\mathcal{F}_*$ denote the set of functions of continuous, bounded functions, 
$$ \mathcal{F}_* := \{ f \in \mathcal{C}(\RR^{3n},\RR) \big| \avg{f}_{\xb} = \avg{f(\xb) \delta(\Theta(\xb) - \zb)}_{\xb, \zb} \},$$
where $\avg{\cdot}_{\xb}$ denotes an ensemble average with respect to the fine-grained Boltzmann distribution $\rho(\xb) d\xb = e^{-\beta U(\xb)} d\xb$ and $\avg{\cdot}_{\xb,\zb}$ is also integrated over $\zb$. 
If $(\Theta, \hat U, T)$ is $\mathcal{F}_*$ thermodynamically consistent, then the projective coarse-graining $(\Theta, \hat U)$ is thermodynamically consistent in the sense of Ref.~\cite{noid_multiscale_2008}.
\end{proposition}
At a high level, this statement says that weak thermodynamic consistency for all observables with average values that are preserved by the projection of the Boltzmann distribution onto the coarse-grained space implies thermodynamic consistency.


\begin{figure}
\centering
    \includegraphics[width=0.7\linewidth]{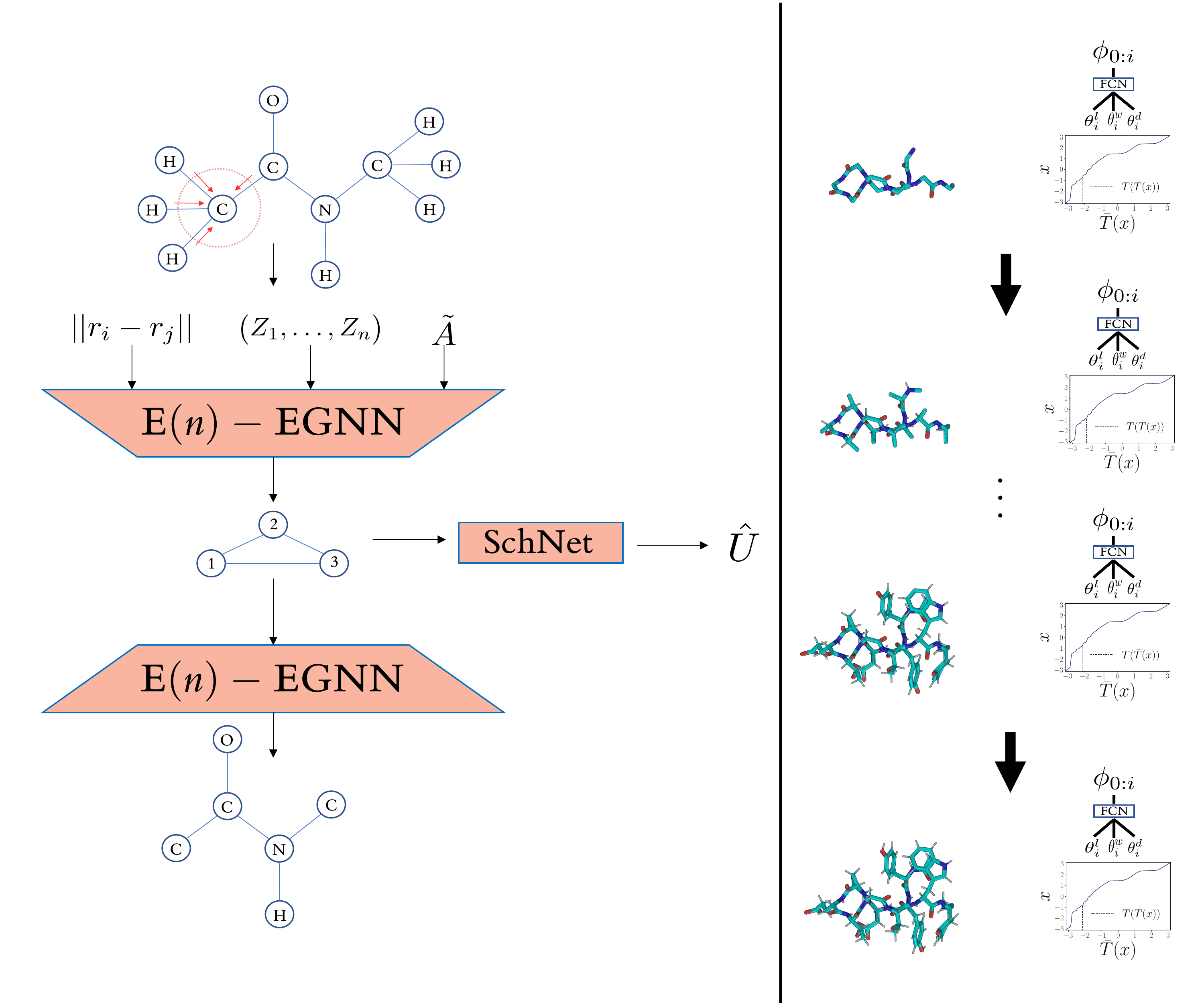}
    \caption{Schematic of computational pipeline for coarse-graining (left) and back-mapping (right). Left. Configurations are shown as 2-D graphs for ease of visualization. Initial configuration $\xb_i$ is passed through coarse-graining network $\Theta$ to determine corresponding coarse-grained configuration $\zb_i$. Coarse-Grained configuration is passed through decoding network $\Theta^{\rm dec}$ to reconstruct a set of target atoms $\tilde{\xb}_i$. Coarse-Grained potential energy $\hat{U}$ is computed as a function of $\zb_i$. Right. Starting with $\tilde{\xb}_i$, dihedral angles are generated using rational-quadratic neural spline flow (RQ-NSF), which are then used to reconstruct atoms adjacent to atoms in $\tilde{\xb}_i$. This procedure is repeated iteratively until full fine-grained structure is generated.}
    \label{fig:nfschematic}
\end{figure}

\section{State-dependent projection mapping} \label{sec:embed}

Neural networks construct a nonlinear embedding of the input data, which is often viewed as constructing a basis in which regression or classification can be performed easily~\cite{bach_breaking_2017,rotskoff_trainability_2022}.
The fact that learned embeddings provide a highly efficient dimensionality reduction has been exploited in molecular contexts, including for reaction coordinates~\cite{chen_datadriven_2021,smith_multidimensional_2018} and searching through chemical space~\cite{gomez-bombarelli_automatic_2018}. 
Nevertheless, this strength has not been thoroughly explored in the context of coarse-graining, despite the fact using that autoencoders in the context of reaction coordinate discovery emphasizes that machine learning is well-suited to finding low-dimensional representations capable of capturing slow degrees of freedom~\cite{mardt_vampnets_2018,sidky_molecular_2020,ribeiro_reweighted_2018,smith_multidimensional_2018}.
Instead, typically the coarse-graining map is specified at the outset based on physical intuition.

In this work, we learn the coarse-graining map, choosing only the dimensionality of the coarse-grained space.
While we have developed a framework for training a coarse-graining map that is an arbitrary nonlinear function (Appendix~\ref{app:compcg}), all results presented here use a state-dependent linear projection map that is itself a nonlinear function of the atomic positions.
We train the model to find optimal projections using a reconstruction loss scheme described in detail in Appendix~\ref{app:compcg}.
Our training procedure resembles the typical paradigm for autoencoders~\cite{kingma_autoencoding_2013}, adding auxiliary loss functions to regularize the learned representation and favor locality in the coarse-grained mapping.

The embedding we use is depicted schematically in Fig.~\ref{fig:nfschematic}. 
Our approach is based on differentiable pooling (\texttt{DiffPool}), an algorithm developed for parameterizable graph coarsening~\cite{Ying_DiffPool_2018}.
We use a pooling layer that consists of an equivariant graph neural network that outputs a projection matrix $P_{\xb}$ given a fine-grained configuration $\xb$.
Thus, while the projection matrix is a nonlinear function of the input coordinates, the representation in the coarse-grained space is a linear transformation of the input coordinates.
This means that we can train the potential energy function for the coarse-grained model with a typical force matching objective~\cite{husic_coarse_2020}.
In our examples, with the auxiliary loss functions we employ, we find that the coarse-graining map is typically only weakly state dependent, as depicted in the first panel of Fig.~\ref{fig:cg_schematic}.
Admittedly, the representation of the coarse-grained configuration becomes more difficult to interpret due to the complicated relationship between the fine-grained configuration and the resulting projection matrix. 
The loss of interpretability instead motivates us to invert the coarse-graining map directly. 

\section{Inverting the coarse-grained samples}\label{sec:invert}

We often use molecular simulations to investigate properties or observables $f:\RR^{3n} \to \RR$ that \textit{require} atomic resolution.
Such observables cannot be mapped onto a coarse-grained configuration, and finding appropriate proxies in the coarse-grained space for a detailed molecular property is challenging in general---there is not a unique strategy. 
Moreover, the strategy we use to embed molecular configurations is not amenable to a physical interpretation. 
In this work, we carry out a two stage process that allows us to reconstruct fine-grained configurations while also leveraging reduced dimensionality of the coarse-grained system to accelerate exploration of the free energy landscape. 
We first sample collections of coarse-grained configurations in proportion to the coarse-grained Boltzmann probability $\hat \rho(\zb) d\zb.$
We subsequently employ an invertible neural network, known as a normalizing flow, to harvest a statistical sample of fine-grained configurations in such a way that we can reweight fine-grained configurations in proportion to their true Boltzmann weight. 

It requires care to ensure that the sampling and subsequent reconstruction can be combined to obtain the correct statistics for the target Boltzmann distribution of the fine-grained system. 
In general, there are two options: Metropolis Monte Carlo or reweighting. 
Sampling the coarse-grained system with the Metropolis-adjusted Langevin Algorithm (MALA) leads to a collection of coarse-grained configurations $\{ \zb_i \}_{i=1}^k$ which are distributed in proportion to the Boltzmann distribution associated with the coarse-grained energy function, that is, $\hat p(z) \propto e^{-\beta \hat{U}(z)}.$
We use the conventional force-matching paradigm to train $\hat U$, for which we use a standard implementation of the SchNet architecture~\cite{schutt_schnet_2018}.
We train the energy function so that it reconstructs the potential of mean-force associated with the fine-grained energy $U$.
While the coarse-grained energy function $\hat{U}$ is often represented with an empirical potential functional form, recently more general functional forms have been employed, using traditional empirical potentials but also adding a general neural network~\cite{wang_machine_2019, husic_coarse_2020}.
Because the optimization of the force-matching objective relies on data collected from fine-grained molecular simulations, the coarse-grained potential will not exactly match the potential of mean force $\hat F.$

Nevertheless, this discrepancy can be systematically corrected. 
We first invert the coarse-grained representation by mapping the coarse-grained configuration via a linear back-projection decoding map $\Theta^{\rm dec}:\RR^{3k} \to \RR^{3\tilde{n}}.$ 
In our examples, the backbone is reconstructed from this map.
We then conditionally sample fine-grained configurations $\xb$ using a normalizing flow $T$, which parameterizes a conditional distribution $p(\xb | \zb)$.
In order to appropriately reweight the samples generated by this procedure, we need to compute the resulting marginal distribution that arises from integrating out the coarse-grained distribution. 
Viewed as an MCMC algorithm, the generation probabilities are explicit functions of the generated configuration and do not depend on the previous configuration within the Markov chain.
We can write
\begin{equation}
    p_{\rm gen}(\xb_{i+1}|\xb_i) = \hat p(\zb_{i+1}) p(\xb_{i+1} | \zb_{i+1})
\end{equation}
where the conditional distribution can be computed using the pushforward probability density. 
The pushforward density is given by inverting the map and evaluating its probability in the Gaussian density, 
\begin{equation}
    T\sharp \varrho(\xb) = \varrho(\bar T(\xb)) | \nabla \bar T(\xb) |
\end{equation}
where $| \nabla \bar T (\xb)|$ denotes the determinant of the Jacobian of the inverse of the normalizing flow $T$, and $\varrho$ is the density of a Gaussian with mean zero and identity covariance.
The architecture of the neural network $T$ is constructed so that all the terms above are easily computable~\cite{durkan_neural_spline_flows}.

\section{Results}
\subsection*{Alanine Dipeptide}
Alanine dipeptide is a standard benchmark for molecular simulation; its dynamics is well-described by two dihedral angles $\phi$ and $\psi$ \cite{tobias_conformational_1992, montgomerypettitt_potential_1985}. We carried out an MD simulation of alanine dipeptide in explicit solvent for a total duration of over 0.5 $\mu s$. From this trajectory, we generated a dataset by sub-sampling 50000 data points, consisting of positions and forces on all 22 atoms of alanine dipeptide. With this sub-sampling approach, we automatically coarse-grain all water molecules out before training. Using this dataset, we trained a coarse-graining map $\Theta$, a coarse-grained potential energy $\hat{U}$ and a normalizing flow to carry out back-mapping using the procedure described in Section~\ref{sec:invert}, \ref{sec:methods} and Appendix~\ref{app:compcg}, \ref{app:backmapping}.

Our coarse-graining map $\Theta$ projects the 22 atoms from the fine-grained configuration to a coarse-grained configuration consisting of 6 beads. Although we allow for state-dependent embeddings to be learned, we observe that the embeddings are the same across all fine-grained configurations in our dataset. Furthermore, we observe that the learned coarse-graining map projects the 5 backbone heavy atoms and the $C_{\beta}$ onto the coarse-grained space. Importantly, this learned map is consistent with physical intuition and with other coarse-graining works that investigate alanine dipeptide \cite{wang_machine_2019, husic_coarse_2020}.

Using the learned coarse-grained potential $\hat{U}$, we run Langevin dynamics to carry out sampling in the coarse-grained space. We ran 6 trajectories with different initial points and from these trajectories sampled 20000 coarse-grained configurations. Finally, for each of these 20000 configurations, we carry out our back-mapping procedure and generate 200 fine-grained configurations for a total of 400000 configurations generated. Of course, there is no guarantee that the collection of generated configurations are Boltzmann distributed; to ensure this, we carry out a reweighting procedure followed by a single step of overdamped Brownian dynamics (see Appendix \ref{app:reweighting}).

\begin{figure}[H]
    \centering
    \includegraphics[width=1.0\linewidth]{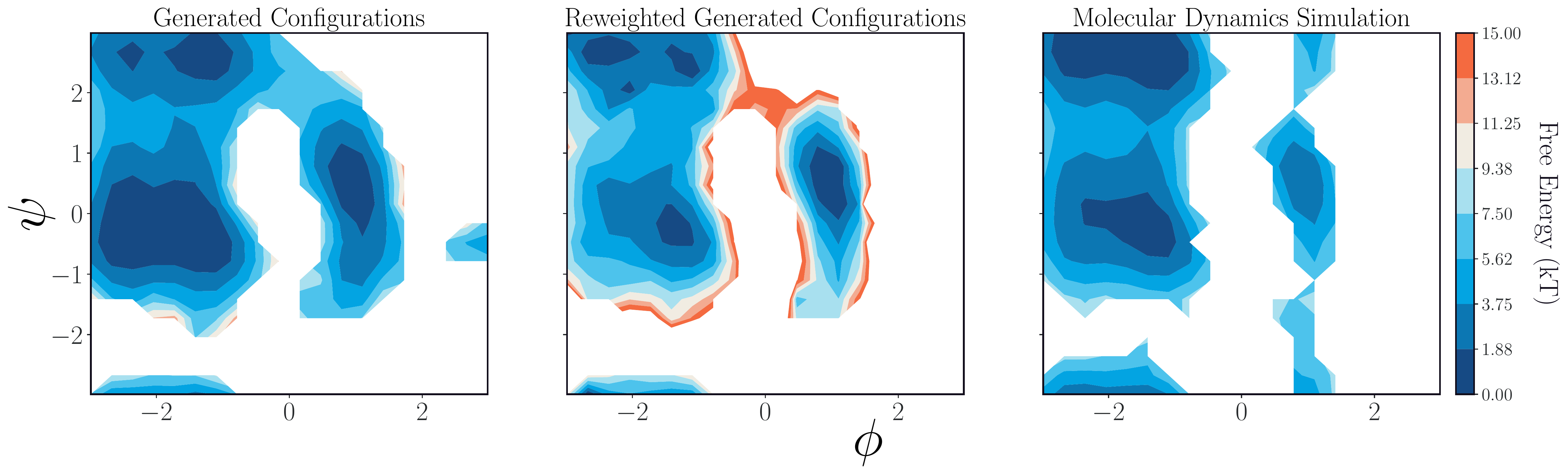}
    \caption{Free-energy landscape of alanine dipeptide as a function of $\phi$ and $\psi$ dihedral angles. Generated configurations from coarse-grained simulations back-mapped into fine-grained configurations (left). Generated configurations from coarse-grained simulations back-mapped into fine-grained configurations with reweighting to ensure configurations are Boltzmann-distributed (center). Training dataset consisting of configurations sampled via MD simulation (right).}
    \label{fig:adp_pmf}
\end{figure}

We plot the free-energy surface as a function of the $\phi$ and $\psi$ dihedral angles in Figure ~\ref{fig:adp_pmf}. Interestingly, we see that in comparison to the free-energy surface obtained from MD simulation, the generated configurations prior to reweighting greatly oversample one of the basins; however, with reweighting the free energy surface of generated configurations closely approximates the free-energy basin obtained via MD simulation and is in strong qualitative agreement with free-energy surfaces documented in literature \cite{deng_connecting_2015}.

\begin{figure}[H]
    \centering
    \includegraphics[width=1.0\linewidth]{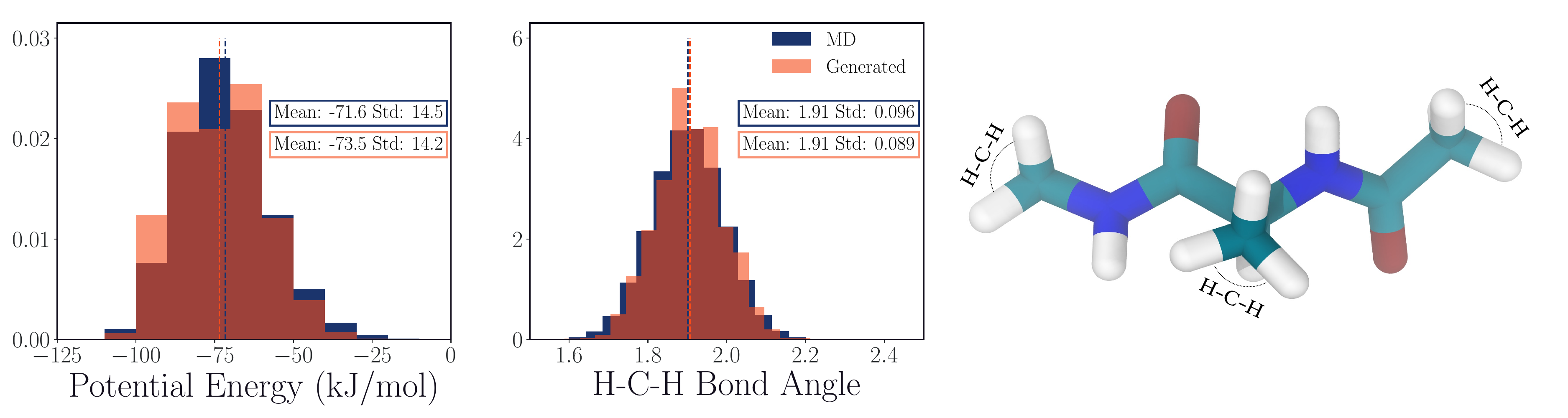}
    \caption{Potential energy of alanine dipeptide computed via an implicit model and H-C-H bond angles between all methyl Hydrogens (right) for reweighted generated configurations (orange) and training dataset obtained via MD (blue). Sample configuration on right illustrates bond angle.}
    \label{fig:adp_dihedrals}
\end{figure}

Finally, we examine two observables that we are unable to compute for a coarse-grained configuration (Figure ~\ref{fig:adp_dihedrals}). First, we look at the potential energy computed via an implicit solvent model for the configurations in our original dataset and our reweighted generated configurations and observe that the mean and variance agree quantitatively. Next, we consider the combined distribution of all 9 H-C-H bond angles between all methyl Hydrogens in alanine dipeptide. We are unable to compute these angles directly from the coarse-grained configurations: back-mapping is essential. Furthermore, these angles are not explicitly used during the back-mapping process. Again, we observe quantitative agreement between the distribution of H-C-H bond angles with our MD dataset and the reweighted generated configurations.

\subsection*{Chignolin}
Next, we carried out a similar analysis for the CLN025 variant of chignolin, a fast-folding micro-protein. Chignolin is a 10-residue protein that adopts a $\beta$-hairpin structure in its folded state \cite{honda_crystal_2008}. Unlike alanine dipeptide, chignolin does not have a set of physical collective variables that can aptly characterize the conformational dynamics. Instead, we utilize time-lagged independent component analysis (tICA) to determine the necessary collective variables \cite{perez-hernandez_identification_2013, schwantes_improvements_2013}. Using a publicly available trajectory of chignolin simulated in explicit solvent \cite{husic_coarse_2020}, we collated a dataset consisting of 50000 positions and forces on all 175 atoms of chignolin. We again coarse-grain out all water molecules before training. As with alanine dipeptide, we train a $\Theta$, $\hat{U}$ and a normalizing flow using the procedure described in Section~\ref{sec:invert}, \ref{sec:methods} and Appendix~\ref{app:compcg}, \ref{app:backmapping}.

For chignolin, the coarse-graining map $\Theta$ projects the 175 atoms in the fine-grained configuration to a coarse-grained configuration consisting of 30 beads. As with alanine dipeptide, we observe that the state-dependent embeddings weakly depend on the fine-grained configurations in our dataset. Furthermore, we observe that the coarse-graining map learns to project backbone atoms $(\rm{C}, \rm{C}_{\alpha}, \rm{N})$ onto the coarse-grained space, again learning an intuitive and physically meaningful map.

With our trained $\hat{U}$, we carried out coarse-grained dynamics consisting of 27 trajectories with different initial points. From these coarse-grained trajectories, we back-mapped 54000 coarse-grained configurations, where we generated 750 fine-grained configurations per coarse-grained configuration. The generated configurations were generally high-energy configurations; however, this was generally a result of minor structural deformities as opposed to major flaws in the reconstruction procedure. To alleviate this, we carry out short overdamped Brownian dynamics in order to relax the structure. Finally, we carried out a reweighting step to ensure that the configurations were Boltzmann-distributed (see Appendix \ref{app:reweighting}).

\begin{figure}[h]
    \centering
    \includegraphics[width=1.0\linewidth]{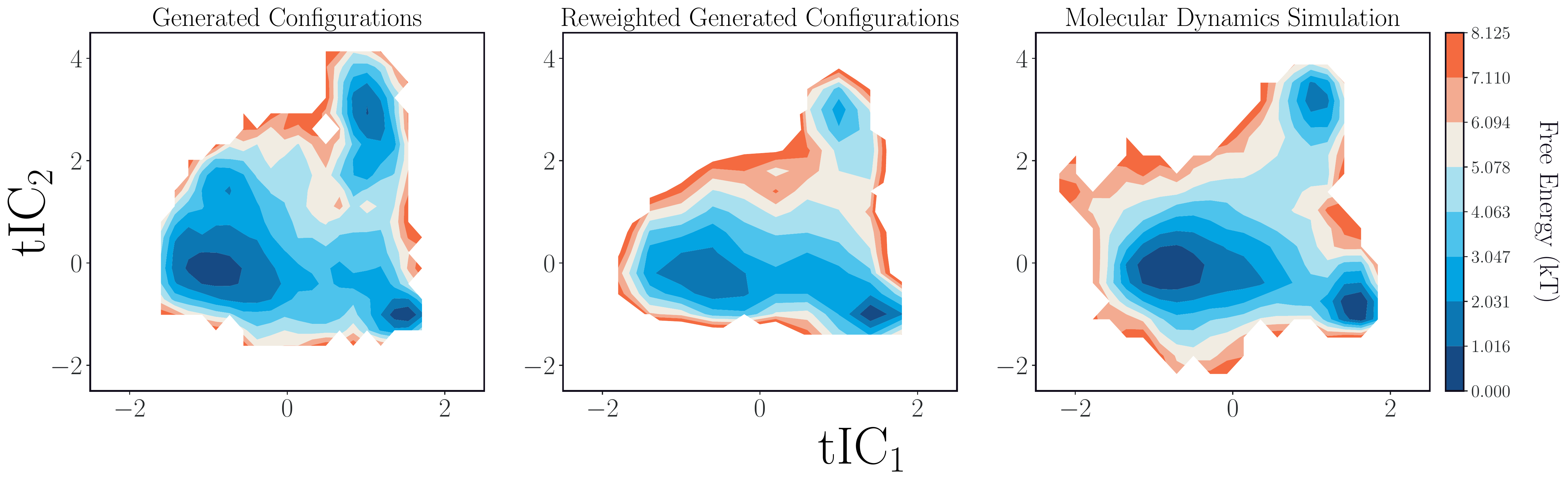}
    \caption{Free-energy landscape of chignolin as a function of Time-Lagged Independent Coordinates ($\rm{tIC}$s). Generated configurations from coarse-grained simulations back-mapped into fine-grained configurations (left). Generated Configurations from coarse-grained simulations back-mapped into fine-grained configurations with reweighting to ensure configurations are Boltzmann-distributed (center). Training dataset consisting of configurations sampled via MD simulation (right).}
    \label{fig:chig_pmf}
\end{figure}

We plot the free-energy surface as a function of the two leading tICA coordinates (see Figure~\ref{fig:chig_pmf}), where we determined the tICA coordinates according to the procedure detailed in \cite{husic_coarse_2020}. The bottom left (first) basin corresponds to the unfolded state, the bottom right (second) basin corresponds to the folded state and the top (third) basin corresponds to the misfolded state. In comparison to the free-energy surface of the dataset obtained via MD simulation, the 
free-energy surface of the generated configurations (prior to reweighting) are heavily populated in regions adjacent to the basins. However, we observe that with reweighting, the free-energy surface of generated configurations more closely matches that of the free-energy surface obtained from MD simulation.

Lastly, we compare two observables that we cannot compute with a coarse-grained configuration. As with alanine dipeptide, we use an implicit solvent model to compute potential energies and observe that the distribution of potential energies from our MD dataset of configurations closely matches the distribution of potential energies of our generated configuration after reweighting. Finally, the hydrophobic side-chains are strongly implicated in the folding of chignolin \cite{mckiernan_modeling_2017}; we probe the rotameric states of the bulky Tryptophan (Trp) residue in chignolin to assess the ability of our back-mapping procedure to faithfully generate the appropriate rotamers. The conformation of the Trp residue is highly dependent on the overall conformation of the protein (i.e. unfolded, folded or misfolded) and with our coarse-grained representation, we are unable to investigate the nature of the rotameric states of the Trp residue. For our reweighted generated structures, we analyze the rotameric state of Tryptophan using the dihedral angle between the $\rm{O}$, $\rm{C}$, $\rm{C}_{\alpha}$ and $\rm{C}_{\beta}$ atoms of the Trp residue. From the MD dataset, it is clear that the distribution of the dihedral angle of the Trp residue is highly dependent on the global conformational state of the protein. We observe that our generated structures are able to closely, but not perfectly, approximate the rotameric distribution of the Trp residue from the MD dataset. Ultimately, we believe that this is a limitation of the normalizing flow architecture used here; we anticipate that with improved normalizing flow models, it will be possible to more closely realize the true distribution of rotameric states.

\begin{figure}[H]
    \centering
    \includegraphics[width=1.0\linewidth]{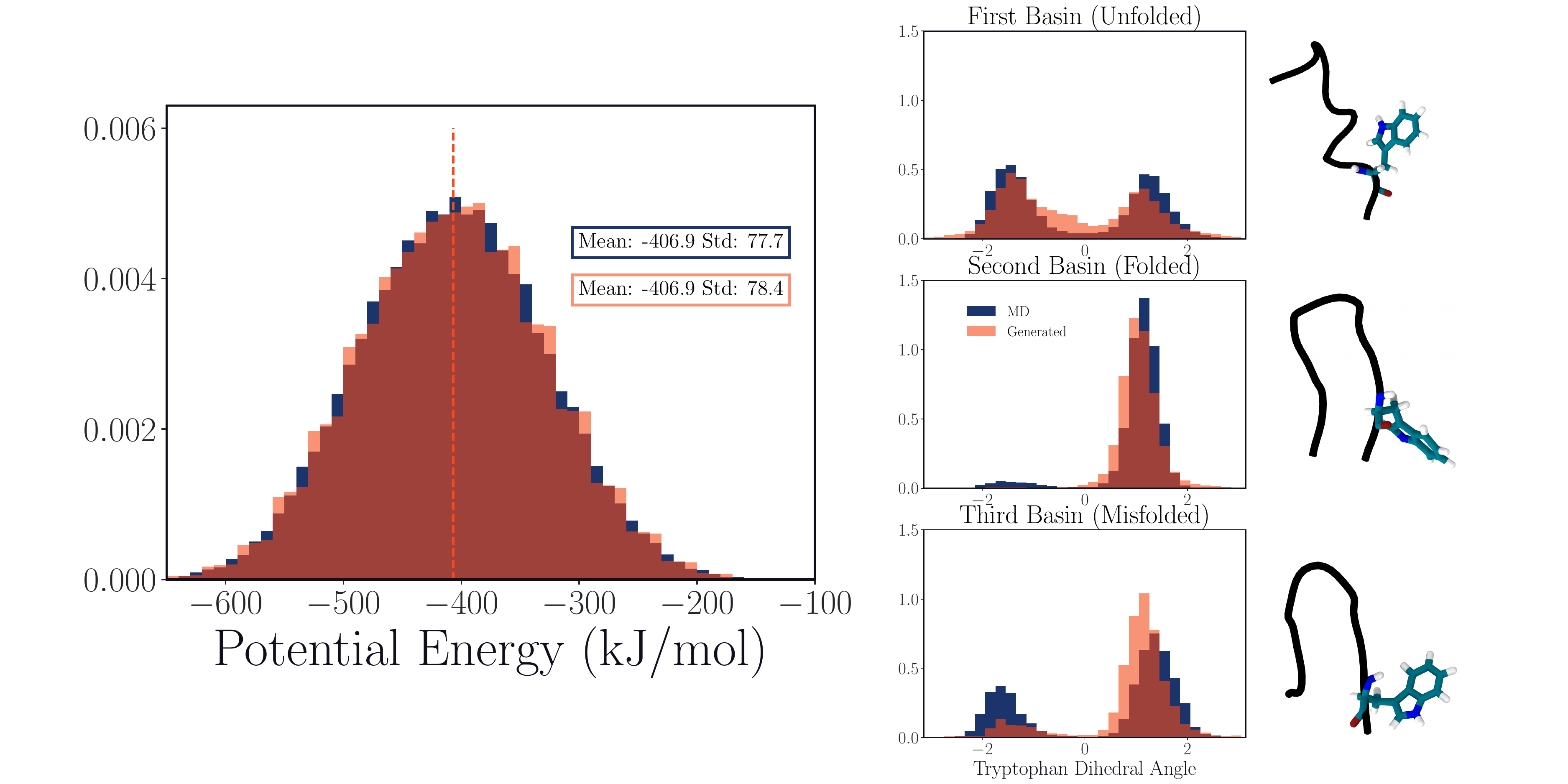}
    \caption{Potential energy of chignolin computed via an implicit model (left) and dihedral angle (between $\rm{O}$, $\rm{C}$, $\rm{C}_{\alpha}$ and $\rm{C}_{\beta}$) of bulky Tryptophan side chain for three different basins of chignolin (right) for reweighted generated configurations (orange) and training dataset obtained via MD (blue). Configuration sampled from reweighted generated configurations for each basin shown, with only backbone and Tryptophan residue shown.}
    \label{fig:chig_dihedrals}
\end{figure}

\section{Methods}\label{sec:methods}
We have developed a computational approach that consists of three interdependent components: a graph neural network representing the coarse-graining map $\Theta: \RR^{3n} \to \RR^{3k}$, a neural network representing the coarse-grained potential energy function $\hat{U}:  \RR^{3k} \to \RR$ and a normalizing flow $T:\RR^{3n}\to \RR^{3n}$ that can generate configurations in the fine-grained space conditioned on a coarse-grained configuration. We work under the assumption that an effective coarse-graining $\Theta$ integrates out degrees of freedom that relax on short timescales, while retaining pertinent information from slower-moving degrees of freedom that inform the free-energy landscape. Additionally, we consider an effective $\Theta$ to be one for which we can train a coarse-grained potential $\hat{U}$ that closely matches the potential of mean force $\hat F$ with a necessarily finite dataset. To account for this multi-task objective, we consider a cyclic training scheme that alternates between training $\Theta$ and $\hat{U}$, where $\Theta$ informs the training of $\hat{U}$ and vice versa. Lastly, we train the normalizing flow independently of the training of $\hat{U}$ and $\Theta$.

We represent our configurations using a three-dimensional graph, where nodes correspond to atoms and edges between nodes correspond to bonded and nonbonded interactions. The graph structure is then coarse-grained through a clustering process, where each cluster corresponds to a coarse-grained ``bead'' and consists of a weighted combination of a collection of nodes. This is achieved via a hierarchical graph pooling technique based on \texttt{DiffPool}\cite{Ying_DiffPool_2018} using the $\rm{E(\emph{n})-EGNN}$ graph neural network architecture, which imposes necessary rotational and physical invariance constraints \cite{satorras_engnn_2021}. 
The resulting embedding is state-dependent; the exact clustering is dependent on the input molecular configuration. We represent our coarse-grained potential energy $\hat{U}$ using the SchNet architecture \cite{schutt_schnet_2018}, which expands inter-bead distances into Gaussian basis functions with learnable parameters; these Gaussian basis functions are then passed through multiple neural network layers. This architecture similarly imposes the necessary physical constraints, most notably rotational and translation invariance.

In order to train $\Theta,$ we consider the encoding-decoding tasks of learning an embedding to determine a coarse-grained configuration $\zb = \Theta(\xb) \in \RR^{3k}$, which can then be used to accurately reconstruct a set of target atoms $\tilde{\xb} = \Theta^{\rm dec}(\zb) \in \RR^{3\tilde{n}}$ from the original fine-grained configuration $\xb\in \RR^{3n}$. The decoding process is achieved via ``inverse pooling'' that is similarly predicated on \texttt{DiffPool}~\cite{Ying_DiffPool_2018} with the $\rm{E(\emph{n})-EGNN}$ architecture~\cite{satorras_engnn_2021}. The target atoms which are reconstructed using the decoder need to be specified \emph{a priori}. In practice, this requires limited knowledge of the system. For the two systems we consider here, we designate the backbone as the set of target atoms to reconstruct.

We first train our coarse-graining network via a reconstruction loss and a suite of auxiliary losses (Appendix~\ref{app:compcg}). The reconstruction loss is the Mean Squared Deviation (MSD) between $\Theta^{\rm dec}(\Theta(\xb)$) and $\tilde{\xb}$. With an initial coarse-graining map, we then train $\hat{U}$ using a mean-force-matching scheme
\begin{equation}
    \mathcal{L}_{\rm{mf}} = [\nabla_{\zb} \hat{U}(\zb) - F_{\rm inst}(\zb)]^2,
    \label{eq:main_mf}
\end{equation}
where $F_{\rm inst}(\zb)$ is the instantaneous mean force, which we use as an estimate of the true mean force (see Appendix~\ref{app:compcg} for details on the mean force). Importantly, $F_{\rm inst}(\zb)$ is a function of $\Theta$, ensuring that $\Theta$ informs the training of $\hat{U}$. 

We can then use $\hat{U}$ to inform the training $\Theta$ using the same objective in Eq.~\ref{eq:main_mf}. In subsequent epochs, we train $\Theta$ using $\mathcal{L}_{\rm{mf}}$ in addition to the reconstruction loss and auxiliary losses. Including $\mathcal{L}_{\rm{mf}}$ penalizes contributions from atoms that contribute to large variances in $F_{\rm inst}(\zb)$, the estimate of the mean force. Physically, these atoms correspond to fast moving degrees-of-freedom. We emphasize that when training $\Theta$, $\hat{U}$ is held fixed, and vice versa.

Finally, from a coarse-grained configuration $\zb$, we would like to generate configurations in the fine-grained space $\xb \sim p(\xb | \zb)$, which we achieve using our decoder $\Theta^{\rm dec}$ and a normalizing flow. To reduce the computational burden of the generative process, we work with internal coordinates: bond angles, bond lengths, and dihedral angles. Furthermore, because the distribution of bond angles and bond lengths are well-approximated by independent Gaussian distributions, with narrow variances, we train a normalizing flow to only sample dihedral angles $\phi \in [-\pi, \pi]^m$ with the bond lengths and bond angles set as the median of their corresponding distributions. We represent our normalizing flow using rational-quadratic neural spline flows (RQ-NSF) with autoregressive layers, primarily because this architecture imposes bounded domains \cite{durkan_neural_spline_flows}. After training our normalizing flow using a forward loss (see Appendix~\ref{app:backmapping}), we can sample dihedral angles $\phi \sim p_{\rm NF}(\phi | \tilde{\xb})$. With $\phi$ and $\tilde{\xb}$, we can easily reconstruct our full fine-grained configuration, a configuration in the fine-grained space (see. Fig.~\ref{fig:nfschematic}). Thus, our back-mapping procedure amounts to sampling $\xb \sim p(\xb | \zb)$.

\section{Discussion}

Coarse-graining has traditionally relied on physical intuition to design effective models for complicated, nonlinear dynamical systems.
With the commensurate loss of intuitive interpretability, we simultaneously exploit the unique capabilities of generative models to reconstruct the atomistic coordinates of the system. 
To interpret convergence, we introduce a conceptual framework for thermodynamic consistency that establishes a notion of equivalence based on averages of observables in the fine-grained space. Additionally, we introduce a highly modular computational pipeline that adheres to this framework. Finally, we demonstrate the capability of our method to quantitatively compute key obervables for two proteins: a simple model system, alanine dipeptide and a fast-folding protein with multiple metastable states, chignolin.

The computational procedure we introduce here is a synthesis of a variety of methods originating from the machine learning and coarse-graining communities.
While practically we were only able to test a small number of different neural network architectures, we believe there is substantial opportunity for improvement as the embedding, the representation of the coarse-grained potential, and the inversion map all become more sophisticated. 
Fortunately, the strategy that we have introduced is highly modular and could even be applied to classical force-matching, projective coarse-graining maps simply by augmenting them with the inversion map. 

In the two systems we investigated here, we had access to a dataset for which the metastable states of interest were well sampled.  
More complex biomolecular systems which are less well characterized will require more sophisticated sampling strategies to generate sufficient data for a generalizable coarse-graining.
We anticipate a feedback approach, where coarse-grained simulations and full molecular dynamic simulations are used in tandem to map out the free energy landscape of the system.

\section*{Acknowledgements}

The authors thank Sherry Li for helpful discussions on Normalizing Flows. This work was supported by the Department of Energy Early Career Research Program under contract DE-SC0022917.

\paragraph*{Data and Code Availability:}
The data that support the findings of this study are available from the corresponding author upon reasonable request. Our code is available at \url{https://github.com/rotskoff-group/thermodynamic-consistency}.

\bibliographystyle{unsrt}
\bibliography{references_shriram,refs}

\begin{thebibliography}{10}

\bibitem{pak_advances_2018}
Alexander~J Pak and Gregory~A Voth.
\newblock Advances in coarse-grained modeling of macromolecular complexes.
\newblock {\em Current Opinion in Structural Biology}, 52:119--126, October
  2018.

\bibitem{noid_multiscale_2008}
W.~G. Noid, Jhih-Wei Chu, Gary~S. Ayton, Vinod Krishna, Sergei Izvekov,
  Gregory~A. Voth, Avisek Das, and Hans~C. Andersen.
\newblock The multiscale coarse-graining method. {{I}}. {{A}} rigorous bridge
  between atomistic and coarse-grained models.
\newblock {\em The Journal of Chemical Physics}, 128(24):244114, June 2008.

\bibitem{clementi_coarsegrained_2008}
Cecilia Clementi.
\newblock Coarse-grained models of protein folding: Toy models or predictive
  tools?
\newblock {\em Current Opinion in Structural Biology}, 18(1):10--15, February
  2008.

\bibitem{saunders_coarsegraining_2012}
Marissa~G Saunders and Gregory~A Voth.
\newblock Coarse-graining of multiprotein assemblies.
\newblock {\em Current Opinion in Structural Biology}, 22(2):144--150, April
  2012.

\bibitem{izvekov_multiscale_2005}
Sergei Izvekov and Gregory~A. Voth.
\newblock A {{Multiscale Coarse-Graining Method}} for {{Biomolecular Systems}}.
\newblock {\em The Journal of Physical Chemistry B}, 109(7):2469--2473,
  February 2005.

\bibitem{izvekov_multiscale_2005a}
Sergei Izvekov and Gregory~A. Voth.
\newblock Multiscale coarse graining of liquid-state systems.
\newblock {\em The Journal of Chemical Physics}, 123(13):134105, October 2005.

\bibitem{wang_coarsegraining_2019}
Wujie Wang and Rafael {G{\'o}mez-Bombarelli}.
\newblock Coarse-graining auto-encoders for molecular dynamics.
\newblock {\em npj Computational Materials}, 5(1):125, December 2019.

\bibitem{wang_machine_2019}
Jiang Wang, Simon Olsson, Christoph Wehmeyer, Adri{\`a} P{\'e}rez, Nicholas~E.
  Charron, Gianni {de Fabritiis}, Frank No{\'e}, and Cecilia Clementi.
\newblock Machine {{Learning}} of {{Coarse-Grained Molecular Dynamics Force
  Fields}}.
\newblock {\em ACS Central Science}, 5(5):755--767, May 2019.

\bibitem{husic_coarse_2020}
Brooke~E. Husic, Nicholas~E. Charron, Dominik Lemm, Jiang Wang, Adri{\`a}
  P{\'e}rez, Maciej Majewski, Andreas Kr{\"a}mer, Yaoyi Chen, Simon Olsson,
  Gianni {de Fabritiis}, Frank No{\'e}, and Cecilia Clementi.
\newblock Coarse graining molecular dynamics with graph neural networks.
\newblock {\em The Journal of Chemical Physics}, 153(19):194101, November 2020.

\bibitem{das_lowdimensional_2006}
Payel Das, Mark Moll, Hern{\'a}n Stamati, Lydia~E. Kavraki, and Cecilia
  Clementi.
\newblock Low-dimensional, free-energy landscapes of protein-folding reactions
  by nonlinear dimensionality reduction.
\newblock {\em Proceedings of the National Academy of Sciences},
  103(26):9885--9890, June 2006.

\bibitem{walther_multimodal_2020}
J{\"u}rgen Walther, Pablo~D Dans, Alexandra Balaceanu, Adam Hospital, Gen{\'i}s
  Bayarri, and Modesto Orozco.
\newblock A multi-modal coarse grained model of {{DNA}} flexibility mappable to
  the atomistic level.
\newblock {\em Nucleic Acids Research}, 48(5):e29, March 2020.

\bibitem{Ying_DiffPool_2018}
Zhitao Ying, Jiaxuan You, Christopher Morris, Xiang Ren, Will Hamilton, and
  Jure Leskovec.
\newblock Hierarchical graph representation learning with differentiable
  pooling.
\newblock In S.~Bengio, H.~Wallach, H.~Larochelle, K.~Grauman,
  N.~{Cesa-Bianchi}, and R.~Garnett, editors, {\em Advances in Neural
  Information Processing Systems}, volume~31. {Curran Associates, Inc.}, 2018.

\bibitem{wang_gen_cg_2021}
Wujie Wang, Minkai Xu, Chen Cai, Benjamin~K Miller, Tess Smidt, Yusu Wang, Jian
  Tang, and Rafael {Gomez-Bombarelli}.
\newblock Generative coarse-graining of molecular conformations.
\newblock In Kamalika Chaudhuri, Stefanie Jegelka, Le~Song, Csaba Szepesvari,
  Gang Niu, and Sivan Sabato, editors, {\em Proceedings of the 39th
  International Conference on Machine Learning}, volume 162 of {\em Proceedings
  of Machine Learning Research}, pages 23213--23236. {PMLR}, July 2022.

\bibitem{kohler_smooth_nfs_2021}
Jonas K{\"o}hler, Andreas Kr{\"a}mer, and Frank Noe.
\newblock Smooth normalizing flows.
\newblock In M.~Ranzato, A.~Beygelzimer, Y.~Dauphin, P.S. Liang, and J.~Wortman
  Vaughan, editors, {\em Advances in Neural Information Processing Systems},
  volume~34, pages 2796--2809. {Curran Associates, Inc.}, 2021.

\bibitem{noe_boltzmann_2019}
Frank No{\'e}, Simon Olsson, Jonas K{\"o}hler, and Hao Wu.
\newblock Boltzmann generators: {{Sampling}} equilibrium states of many-body
  systems with deep learning.
\newblock {\em Science}, 365(6457):eaaw1147, September 2019.

\bibitem{wang_data_2022}
Yihang Wang, Lukas Herron, and Pratyush Tiwary.
\newblock From data to noise to data for mixing physics across temperatures
  with generative artificial intelligence.
\newblock {\em Proceedings of the National Academy of Sciences},
  119(32):e2203656119, August 2022.

\bibitem{kohler_flowmatching_2022}
Jonas K{\"o}hler, Yaoyi Chen, Andreas Kr{\"a}mer, Cecilia Clementi, and Frank
  No{\'e}.
\newblock Flow-matching -- efficient coarse-graining molecular dynamics without
  forces.
\newblock 2022.

\bibitem{mahmoud_accurate_2022}
Amr~H. Mahmoud, Matthew Masters, Soo~Jung Lee, and Markus~A. Lill.
\newblock Accurate {{Sampling}} of {{Macromolecular Conformations Using
  Adaptive Deep Learning}} and {{Coarse-Grained Representation}}.
\newblock {\em Journal of Chemical Information and Modeling}, 62(7):1602--1617,
  April 2022.

\bibitem{tabak_density_2010}
Esteban~G. Tabak and Eric {Vanden-Eijnden}.
\newblock Density estimation by dual ascent of the log-likelihood.
\newblock {\em Communications in Mathematical Sciences}, 8(1):217--233, 2010.

\bibitem{rezende_variational_2015}
Danilo Rezende and Shakir Mohamed.
\newblock Variational {{Inference}} with {{Normalizing Flows}}.
\newblock In {\em International {{Conference}} on {{Machine Learning}}}, pages
  1530--1538. {PMLR}, June 2015.

\bibitem{papamakarios_normalizing_2021}
George Papamakarios, Eric Nalisnick, Danilo~Jimenez Rezende, Shakir Mohamed,
  and Balaji Lakshminarayanan.
\newblock Normalizing flows for probabilistic modeling and inference.
\newblock {\em Journal of Machine Learning Research}, 22(57):1--64, 2021.

\bibitem{albergo_flowbased_2019}
M.~S. Albergo, G.~Kanwar, and P.~E. Shanahan.
\newblock Flow-based generative models for {{Markov}} chain {{Monte Carlo}} in
  lattice field theory.
\newblock {\em Physical Review D}, 100(3):034515, August 2019.

\bibitem{gabrie_adaptive_2022}
Marylou Gabri{\'e}, Grant~M. Rotskoff, and Eric {Vanden-Eijnden}.
\newblock Adaptive {{Monte Carlo}} augmented with normalizing flows.
\newblock {\em Proceedings of the National Academy of Sciences},
  119(10):e2109420119, March 2022.

\bibitem{gabrie_efficient_2021}
Marylou Gabri{\'e}, Grant~M. Rotskoff, and Eric {Vanden-Eijnden}.
\newblock Efficient {{Bayesian Sampling Using Normalizing Flows}} to {{Assist
  Markov Chain Monte Carlo Methods}}.
\newblock In {\em {{ICML Workshop}} on {{Invertible Neural Networks}},
  {{Normalizing Flows}}, and {{Explicit Likelihood Models}}}, 2021.

\bibitem{bach_breaking_2017}
Francis Bach.
\newblock Breaking the {{Curse}} of {{Dimensionality}} with {{Convex Neural
  Networks}}.
\newblock {\em Journal of Machine Learning Research}, 18(19):1--53, 2017.

\bibitem{rotskoff_trainability_2022}
Grant Rotskoff and Eric {Vanden-Eijnden}.
\newblock Trainability and {{Accuracy}} of {{Artificial Neural Networks}}: {{An
  Interacting Particle System Approach}}.
\newblock {\em Communications on Pure and Applied Mathematics},
  75(9):1889--1935, 2022.

\bibitem{chen_datadriven_2021}
Shiqi Chen, Curtis~W. Peterson, John~A. Parker, Stuart~A. Rice, Andrew~L.
  Ferguson, and Norbert~F. Scherer.
\newblock Data-driven reaction coordinate discovery in overdamped and
  non-conservative systems: Application to optical matter structural
  isomerization.
\newblock {\em Nature Communications}, 12(1):2548, May 2021.

\bibitem{smith_multidimensional_2018}
Zachary Smith, Debabrata Pramanik, Sun-Ting Tsai, and Pratyush Tiwary.
\newblock Multi-dimensional spectral gap optimization of order parameters
  ({{SGOOP}}) through conditional probability factorization.
\newblock {\em The Journal of Chemical Physics}, 149(23):234105, December 2018.

\bibitem{gomez-bombarelli_automatic_2018}
Rafael {G{\'o}mez-Bombarelli}, Jennifer~N. Wei, David Duvenaud, Jos{\'e}~Miguel
  {Hern{\'a}ndez-Lobato}, Benjam{\'i}n {S{\'a}nchez-Lengeling}, Dennis
  Sheberla, Jorge {Aguilera-Iparraguirre}, Timothy~D. Hirzel, Ryan~P. Adams,
  and Al{\'a}n {Aspuru-Guzik}.
\newblock Automatic {{Chemical Design Using}} a {{Data-Driven Continuous
  Representation}} of {{Molecules}}.
\newblock {\em ACS Central Science}, 4(2):268--276, February 2018.

\bibitem{mardt_vampnets_2018}
Andreas Mardt, Luca Pasquali, Hao Wu, and Frank No{\'e}.
\newblock {{VAMPnets}} for deep learning of molecular kinetics.
\newblock {\em Nature Communications}, 9(1):5, January 2018.

\bibitem{sidky_molecular_2020}
Hythem Sidky, Wei Chen, and Andrew~L. Ferguson.
\newblock Molecular latent space simulators.
\newblock {\em Chemical Science}, 11(35):9459--9467, 2020.

\bibitem{ribeiro_reweighted_2018}
Jo{\~a}o Marcelo~Lamim Ribeiro, Pablo Bravo, Yihang Wang, and Pratyush Tiwary.
\newblock Reweighted autoencoded variational {{Bayes}} for enhanced sampling
  ({{RAVE}}).
\newblock {\em The Journal of Chemical Physics}, 149(7):072301, August 2018.

\bibitem{kingma_autoencoding_2013}
Diederik~P. Kingma and Max Welling.
\newblock Auto-{{Encoding Variational Bayes}}.
\newblock December 2013.

\bibitem{schutt_schnet_2018}
K.~T. Sch{\"u}tt, H.~E. Sauceda, P.-J. Kindermans, A.~Tkatchenko, and K.-R.
  M{\"u}ller.
\newblock {{SchNet}} \textendash{} {{A}} deep learning architecture for
  molecules and materials.
\newblock {\em The Journal of Chemical Physics}, 148(24):241722, June 2018.

\bibitem{durkan_neural_spline_flows}
Conor Durkan, Artur Bekasov, Iain Murray, and George Papamakarios.
\newblock Neural spline flows.
\newblock In H.~Wallach, H.~Larochelle, A.~Beygelzimer, F.~{dAlch{\'e}-Buc},
  E.~Fox, and R.~Garnett, editors, {\em Advances in Neural Information
  Processing Systems}, volume~32. {Curran Associates, Inc.}, 2019.

\bibitem{tobias_conformational_1992}
Douglas~J. Tobias and Charles~L. Brooks.
\newblock Conformational equilibrium in the alanine dipeptide in the gas phase
  and aqueous solution: A comparison of theoretical results.
\newblock {\em The Journal of Physical Chemistry}, 96(9):3864--3870, April
  1992.

\bibitem{montgomerypettitt_potential_1985}
B.~Montgomery~Pettitt and Martin Karplus.
\newblock The potential of mean force surface for the alanine dipeptide in
  aqueous solution: A theoretical approach.
\newblock {\em Chemical Physics Letters}, 121(3):194--201, November 1985.

\bibitem{deng_connecting_2015}
Nanjie Deng, Bin~W. Zhang, and Ronald~M. Levy.
\newblock Connecting {{Free Energy Surfaces}} in {{Implicit}} and {{Explicit
  Solvent}}: {{An Efficient Method To Compute Conformational}} and {{Solvation
  Free Energies}}.
\newblock {\em Journal of Chemical Theory and Computation}, 11(6):2868--2878,
  June 2015.

\bibitem{honda_crystal_2008}
Shinya Honda, Toshihiko Akiba, Yusuke~S. Kato, Yoshito Sawada, Masakazu
  Sekijima, Miyuki Ishimura, Ayako Ooishi, Hideki Watanabe, Takayuki Odahara,
  and Kazuaki Harata.
\newblock Crystal {{Structure}} of a {{Ten-Amino Acid Protein}}.
\newblock {\em Journal of the American Chemical Society}, 130(46):15327--15331,
  November 2008.

\bibitem{perez-hernandez_identification_2013}
Guillermo {P{\'e}rez-Hern{\'a}ndez}, Fabian Paul, Toni Giorgino, Gianni
  De~Fabritiis, and Frank No{\'e}.
\newblock Identification of slow molecular order parameters for {{Markov}}
  model construction.
\newblock {\em The Journal of Chemical Physics}, 139(1):015102, July 2013.

\bibitem{schwantes_improvements_2013}
Christian~R. Schwantes and Vijay~S. Pande.
\newblock Improvements in {{Markov State Model Construction Reveal Many
  Non-Native Interactions}} in the {{Folding}} of {{NTL9}}.
\newblock {\em Journal of Chemical Theory and Computation}, 9(4):2000--2009,
  April 2013.

\bibitem{mckiernan_modeling_2017}
Keri~A. McKiernan, Brooke~E. Husic, and Vijay~S. Pande.
\newblock Modeling the mechanism of {{CLN025}} beta-hairpin formation.
\newblock {\em The Journal of Chemical Physics}, 147(10):104107, September
  2017.

\bibitem{satorras_engnn_2021}
Victor~Garcia Satorras, Emiel Hoogeboom, and Max Welling.
\newblock E(n) equivariant graph neural networks.
\newblock In Marina Meila and Tong Zhang, editors, {\em Proceedings of the 38th
  International Conference on Machine Learning, {{ICML}} 2021, 18-24 July 2021,
  Virtual Event}, volume 139 of {\em Proceedings of Machine Learning Research},
  pages 9323--9332. {PMLR}, 2021.

\bibitem{ciccotti_projection_2008}
Giovanni Ciccotti, Tony Leli{\`e}vre, and Eric {Vanden-Eijnden}.
\newblock Projection of diffusions on submanifolds: {{Application}} to mean
  force computation.
\newblock {\em Communications on Pure and Applied Mathematics}, 61(3):371--408,
  March 2008.

\bibitem{kalligiannaki_geometry_2015}
Evangelia Kalligiannaki, Vagelis Harmandaris, Markos~A. Katsoulakis, and Petr
  Plech{\'a}{\v c}.
\newblock The geometry of generalized force matching and related information
  metrics in coarse-graining of molecular systems.
\newblock {\em The Journal of Chemical Physics}, 143(8):084105, August 2015.

\bibitem{sivak_time_2014}
David~A. Sivak, John~D. Chodera, and Gavin~E. Crooks.
\newblock Time {{Step Rescaling Recovers Continuous-Time Dynamical Properties}}
  for {{Discrete-Time Langevin Integration}} of {{Nonequilibrium Systems}}.
\newblock {\em The Journal of Physical Chemistry B}, 118(24):6466--6474, June
  2014.

\bibitem{eastman_openmm_2017}
Peter Eastman, Jason Swails, John~D. Chodera, Robert~T. McGibbon, Yutong Zhao,
  Kyle~A. Beauchamp, Lee-Ping Wang, Andrew~C. Simmonett, Matthew~P. Harrigan,
  Chaya~D. Stern, Rafal~P. Wiewiora, Bernard~R. Brooks, and Vijay~S. Pande.
\newblock {{OpenMM}} 7: {{Rapid}} development of high performance algorithms
  for molecular dynamics.
\newblock {\em PLOS Computational Biology}, 13(7):e1005659, July 2017.

\end{thebibliography}

\appendix

\newpage
\section{Computational Details for Coarse-Graining}\label{app:compcg}
We use a training scheme that alternates between training a coarse-graining function $\Theta$ and the coarse-grained energy function $\hat{U}$. In this scheme, we learn an initial $\Theta_0$, which is then used to learn an initial $\hat{U}_0$. This $\hat{U}_0$ is then used to partially inform the training of a new $\Theta_1$, with this feedback approach continuing for a predefined number of epochs $\nb_{\rm freeze\_cg}$. After $\nb_{\rm freeze\_cg}$ training epochs, we fix the coarse-graining function $\Theta$ and train $\hat{U}$ until convergence.

We represent our configurations using a three-dimensional graph, where nodes correspond to atoms and edges between nodes correspond to bonded and nonbonded interactions. For a given atom, atoms within a prespecified nonbonded edge cutoff are connected via a nonbonded edge if a bonded edge does not already exist. We use atomic numbers as input node features and the type of the edge (bonded, nonbonded, self) as the input edge feature.

To learn a coarse-graining function $\Theta,$ we consider the encoding-decoding tasks of learning an embedding to determine a coarse-grained configuration $\zb_i = \Theta(\xb_i) \in \RR^{3k}$, which can then be used to accurately reconstruct a set of target atoms $\tilde{\xb}_i = \Theta^{\rm dec}(\zb_i) \in \RR^{3\tilde{n}}$ from the original fine-grained configuration $x_i$ $\in \RR^{3n}$. The target atoms which are reconstructed using the decoder need to be specified \emph{a priori}. In practice, this requires limited knowledge of the system. For the two model systems we consider here, we designate the backbone as the set of target atoms to reconstruct.

We model our encoder-decoder (Fig.~\ref{fig:nfschematic}) scheme after $\texttt{DiffPool}$, a hierarchical graph pooling approach \cite{Ying_DiffPool_2018}. Our encoder consists of a single pooling layer, where we learn a fine-grained-configuration-dependent projection matrix $P_{\xb_i}$, such that $\zb_i = \Theta(\xb_i) = P_{\xb_i}\xb_i$. Similarly, the decoder consists of a single ``inverse-pooling'' layer, where we learn a coarse-grained-configuration-dependent projection matrix, $P_{\zb_i}$ in order to reconstruct $\bar{\xb}_i = \Theta^{\rm dec}(\zb_i) = P_{\zb_i}\zb_i$. Importantly, the encoder-decoder is state-dependent, ensuring it is consistent with the coarse-graining framework we introduce in the main text. 

Here, $P_{\xb_i} = \rm{softmax}(\rm{GNN}(\xb_i)) \in \RR^{{k}\times{n}}$ and $P_{\zb_i} = \rm{GNN}(\xb_i) \in \RR^{{\tilde{n}}\times{k}}$, where the softmax is computed row-wise to ensure that the sum of all atomic contributions to each coarse-grained ``bead'' is 1. The $\texttt{DiffPool}$ scheme allows for any general message-passing GNN architecture to be used; we use the $\rm{E(\emph{n})-EGNN}$ graph neural network architecture, which imposes necessary rotational and physical invariance constraints \cite{satorras_engnn_2021}.

Our loss function to train this encoder-decoder consists of a reconstruction loss and a suite of auxiliary losses, which aid in regularization
\begin{equation}
    \mathcal{L}_{\Theta} = \mathcal{L}_{\rm{r}} + \lambda [\mathcal{L}_{\rm{link}} + \mathcal{L}_{\rm{ent}} + \mathcal{L}_{\rm{assgn}} + \lambda_{\rm mf}\mathcal{L}_{\rm{mf}}],
\end{equation} where $\lambda$ and $\lambda_{\rm{mf}}$ are hyperparameters controlling the weight of the four auxiliary losses. 
The reconstruction loss $\mathcal{L}_{\rm{r}} = \left\|P_{\zb} P_{\xb} \xb -\tilde{\xb}\right\|_{2}^2$, where $P_{\xb}$ and $P_{\zb}$ are dependent on the initial fine-grained configuration $x$ and the coarse-grained configuration $\zb = P_{\xb} \xb$ respectively. In our implementation, we remove translational shifts before computing the distance. Additionally, when computing $\mathcal{L}_{\rm r}$, we weigh all backbone Carbon and Nitrogen atoms by $\alpha_{\rm bb} > 1$ to ensure that these atoms get reconstructed with a higher fidelity.

The link loss $\mathcal{L}_{\textrm{link}} = \left\| D \odot P_{\xb}^{T}P_{\xb} \right\|_{F}$ ensures that proximal atoms are projected onto the same coarse-grained bead. Here, $D \in \RR^{n\times n}$ is the matrix of all pairwise distances between atoms of the fine-grained configuration $x$. The entropy loss $\mathcal{L}_{\textrm{ent}} = \frac{1}{k}\sum_{i=1}^{k} H(P_{\xb}^i)$, where $P_{\xb}^i$ the $i$-th row of the projection matrix $P_{\xb}$ and $H$ denotes the entropy function. This loss ensures that the fractional weights of the atoms assigned to each bead are concentrated around a few atoms. The assignment loss $\mathcal{L}_{\textrm{assgn}} = \textrm{diag}(P_{\xb} P_{\xb}^T)$ ensures that an atom is not assigned to multiple beads, where $\textrm{diag}$ corresponds to the diagonal entries of the matrix. Lastly, we elaborate on the mean-force loss $\mathcal{L}_{\rm{mf}}$ below after a brief discussion on the computation of the mean-force.

For the systems we explore here, we cannot easily compute a coarse-grained potential energy function $\hat{U}$; instead, we learn $\hat{U}$ using force-matching \cite{izvekov_multiscale_2005, izvekov_multiscale_2005a}. We match $\grad_{\zb}\hat{U}$ to the mean force of a coarse-grained configuration. The mean force formally is
\begin{equation}
    F(z) = \langle \Theta_{F}(\nabla_{\xb} U(\xb)) \rangle_{\Theta(\xb) = \zb},
\end{equation}
where $\Theta_{F}$ projects the fine-grained force $\nabla_{\xb} U(\xb)$ into the coarse-grained space \cite{noid_multiscale_2008}. Computing this average exactly can be costly; instead, we compute an estimate of the mean force using the instantaneous coarse-grained force \cite{ciccotti_projection_2008, kalligiannaki_geometry_2015}
\begin{equation}
    F_{\rm inst}(\zb) = (P_{\xb} P_{\xb}^T)^{-1}P_{\xb}\nabla_{\xb} U(\xb),
    \label{eq:mf}
\end{equation}
where $\zb = \Theta(\xb) = P_{\xb} \xb$.
With this estimate, we train $\hat{U}$ by minimizing the following loss.
\begin{equation}
    \mathcal{L}_{\hat{U}} = [\boldsymbol{\nabla_{\zb} \hat{U}(\zb)} - F_{\rm inst}(\zb)]^2.
    \label{eq:cgu}
\end{equation}
We use the SchNet architecture to represent $\hat{U}$ \cite{schutt_schnet_2018}. Briefly, SchNet transforms interbead interbead into gaussian basis functions with learnable parameters; these gaussian basis functions are then passed through multiple neural network layers. We assign each bead a unique input feature ($0\dots k$), analogous to an atomic number.

Finally, we use our coarse-grained potential energy $\hat{U}$ to inform the coarse-grained map $\Theta$ that we learn using $\mathcal{L}_{\rm{mf}}$
\begin{equation}
    \mathcal{L}_{\rm{mf}} = [\nabla_{\zb} \hat{U}(\zb) - \boldsymbol{F_{\rm inst}(\zb)}]^2.
    \label{eq:cgmf}
\end{equation}
While, $\mathcal{L}_{\rm{mf}}$ has the same form as $\mathcal{L}_{\hat{U}}$, in Eq.~\ref{eq:cgu} we fix our coarse-graining map and allow $\hat{U}$ to train. In Eq.~\ref{eq:cgmf} on the other hand, we fix the coarse-grained potential energy function $\hat{U}$ and instead allow our coarse-graining map to vary. Using $\hat{U}$ to inform $\Theta$ via $\mathcal{L}_{\rm{mf}}$ reduces contributions from atoms that increase the variance of the mean force estimate $F_{\rm inst}(\zb)$, an observation also made using a similar regularization term in \cite{wang_coarsegraining_2019}. Empirically, we observe that this results in a lower entropy projection map. For the experiments included here, we are able to learn a low-entropy map during the first epoch of training $\Theta$, so the inclusion of $\mathcal{L}_{\rm{mf}}$ has limited practical utility for the systems considered here.

\begin{algorithm}
\caption{Encoder-Decoder Training}
\label{alg:cg}
\begin{algorithmic}[1]
    \State {Initialize Encoder $\Theta$ and Decoder $\Theta^{\rm dec}$ and corresponding optimizer and scheduler}
    \For {e = $0$ ${\dots}$ $\Mb$}
        \If {e ${< \nb_{\rm freeze\_cg}}$}
            \For {t = $0$ ${\dots}$ $\nb_{\rm cg}$} 
                \State{Compute $\mathcal{L}_{\Theta}$ and carry out optimization step of $\Theta$}
            \EndFor
            \State{(Re)-Initialize $\hat{U}$ and corresponding optimizer and scheduler}
        \EndIf
        \For {t = $0$ ${\dots}$ $\nb_{\rm u}$}
            \State{Compute $\mathcal{L}_{\hat{U}}$ and carry out optimization step of $\hat{U}$}
        \EndFor
    \EndFor
\end{algorithmic}
\end{algorithm}

\section{Computational Details for back-mapping}\label{app:backmapping}
In the coarse-graining scheme discussed in Appendix~\ref{app:compcg}, we train an encoder-decoder that deterministically embeds a fine-grained configuration $x_i$ into a coarse-grained configuration $\zb_i$; the coarse-grained configuration is then used to deterministically reconstruct $\tilde{\xb}_i$, which corresponds to a set of target atoms from the original configuration. Of course, we cannot compute observables dependent on the full fine-grained configuration using this approach. Here, we describe the back-mapping procedure used to generate fine-grained configurations from a coarse-grained configuration $p(\xb | \zb_i)$. 

To simplify the back-mapping process, we work with internal coordinates, namely bond lengths, bond angles and dihedral angles. From a set of internal coordinates, the Cartesian coordinates of each atom in the system can be easily computed. We use a normalizing flow to generate internal coordinates conditioned on a coarse-grained configuration.  Normalizing flows are a class of invertible neural networks that enable a transformation between two distributions. Importantly, normalizing flows enable efficient and exact density estimation \cite{gabrie_adaptive_2022}. 

For the systems we consider here, the distributions of bond angles and distances are unimodal Gaussian distributions with narrow variances. On the other hand, the distributions of dihedral angles are generally multimodal with large variances. To simplify the generative process, we set all bond angles and distances to be the median of their respective distributions. Thus, we only consider the task of generating dihedral angles
$\phi_i = [\phi_i^0, \phi_i^1 \dots \phi_i^{m}] \in [-\pi, \pi]^m$.

To carry out the back-mapping process, we first use the decoder $\Theta^{\rm dec}$ to transform $\zb_i$ into $\tilde{\xb}_i$. We then compute $\phi^{\rm seed} \in [-\pi, \pi]^{\tilde{m}}$, the dihedral angles of $\tilde{\xb}_i$. Finally, we use our trained normalizing flow to sample a set of dihedral angles $\phi_i \sim p(\phi | \phi^{\rm seed})$. Using $\tilde{\xb}_i$ and $\phi_i$, we can easily generate $\xb_i$, a configuration in the fine-grained space. Thus, our back-mapping procedure amounts to sampling $\xb_i \sim p(\xb | \zb_i)$.

In order to train our normalizing flow, we seek to learn a map $T_*$ that transports the base distribution $\varrho$, a multivariate Gaussian distribution with mean zero and identity covariance, to the target distribution $\tilde{\rho}$, the distribution of internal coordinates that correspond to Boltzmann-distributed configurations. Denoting the map represented by normalizing flow $T$,
\begin{equation}
    T\sharp \varrho(\phi) = \varrho(\bar{T}(\phi)) \left|\grad_\phi \bar{T}\right|,
    \label{eq:rhohat}
\end{equation}
where $\bar{T}$ denotes the inverse map. In order to learn a map $T$ that approximates $T_*$, we seek to minimize the forward Kullback-Leibler Divergence, $D_{\rm KL}(\tilde{\rho} | T\sharp \varrho\ )$. The KL divergence, up to a constant, can be estimated using the following loss function
\begin{equation}
    \mathcal{L}_{\rm NF} = -\frac{1}{N}\sum_{i=1}^{N} \log T\sharp \varrho({\phi_i}),
\end{equation}
where the exact density $T\sharp \varrho$ can be computed using Eq.~\ref{eq:rhohat} and $N$ is the number of data points.

We use the rational-quadratic neural spline flow (RQ-NSF) architecture to represent our normalizing flow \cite{durkan_neural_spline_flows}. We make this choice because the domain of dihedral angles is bounded to $[-\pi, \pi]$ and the RQ-NSF approach considers transformations between bounded domains. Finally, within this architecture we use autoregressive layers to carry out the actual normalizing flow. Practically, this amounts to using the first $j$ dihedral angles, $\phi^{0:j-1}$, to inform the generative process of $\phi^j$.

With RQ-NSF, we define a set of $m$ Neural Networks ${\rm FCN}^0, {\rm FCN}^1, \dots {\rm FCN}^m$ that are used in the transformation of $\phi^0, \phi^1, \dots \phi^m$, respectively. Each of these neural networks are fully connected with a single hidden layer. Given a sample from the base distribution $\phi_{b_i} \sim \varrho_b$, we can compute $\theta_i^j = {\rm FCN}^j(\phi_i^{0:j-1}, \phi^{\rm seed})$.
Here, $\theta_i^j$ correspond to the parameters of a rational quadratic spline $g$. Finally, we can compute $\phi_i^j = g_{\theta_i^j}(\phi_{b_i}^j)$. See Algorithm~\ref{alg:nf} for a summary of the back-mapping process. For a thorough description of RQ-NSF, see \cite{durkan_neural_spline_flows}.

There is a natural hierarchy to the back-mapping procedure here, where atoms that are a single bond away from the atoms in $\tilde{x}$ are reconstructed first with atoms further away being reconstructed later. When carrying out the autoregressive flow, we remain faithful to this hierarchy, ensuring that atoms reconstructed first influence the internal coordinates of atoms reconstructed later.

\begin{algorithm}
\caption{Sampling $\xb_i$ from $p(\xb | \zb_i)$}
\label{alg:nf}
\begin{algorithmic}[1]
    \State {Train ${\rm FCN}^0, {\rm FCN}^1, \dots {\rm FCN}^m$}
    \State {Compute $\phi^{\rm seed}$ from $\tilde{\xb}_i = \Theta^{\rm dec} (\zb_i)$}
    \State {Sample $\phi_{b_i} \sim \varrho$}
    \For {$j = 0 {\dots}$ $m$}
        \State{Compute $\theta_i^j = \rm{FCN}(\phi_i^{0:j-1}, \phi^{\rm seed})$}
        \State{Compute $\phi_i^j = g_{\theta_i^j}(\phi_{b_i}^j)$}
    \EndFor
    \State {Reconstruct $\xb_i$ from $\tilde{\xb}_i$ and $\phi_i$}
\end{algorithmic}
\end{algorithm}

\section{Numerical Experiments}
As described in the main text, we carry out our coarse-graining and back-mapping procedures for two proteins: Alanine Dipeptide and Chignolin. Here, we briefly expand on some of the experimental details used when investigating these systems, including a list of hyperparameters used (see Table~\ref{tab:cgu} and Table~\ref{tab:nf}). 

During training, we use the ReduceLROnPlateau scheduler to train $\Theta$ and $\hat{U}$. The scheduler is assigned a metric of interest and it anneals the learning rate as this metric converges. For $\Theta,$ we use the reconstruction loss $\mathcal{L}_{\rm{r}}$ on the validation set as the metric for the scheduler. For $\hat{U},$ we use the mean absolute error (MAE) between $\nabla \hat{U}(z)$ and $\boldsymbol{F_{\rm inst}(z)}$ on the validation set as the metric for the scheduler. For both Chignolin and Alanine Dipeptide, we use a 80/20 train-validation split.
\begin{figure}
    \centering
    \includegraphics[width=\linewidth]{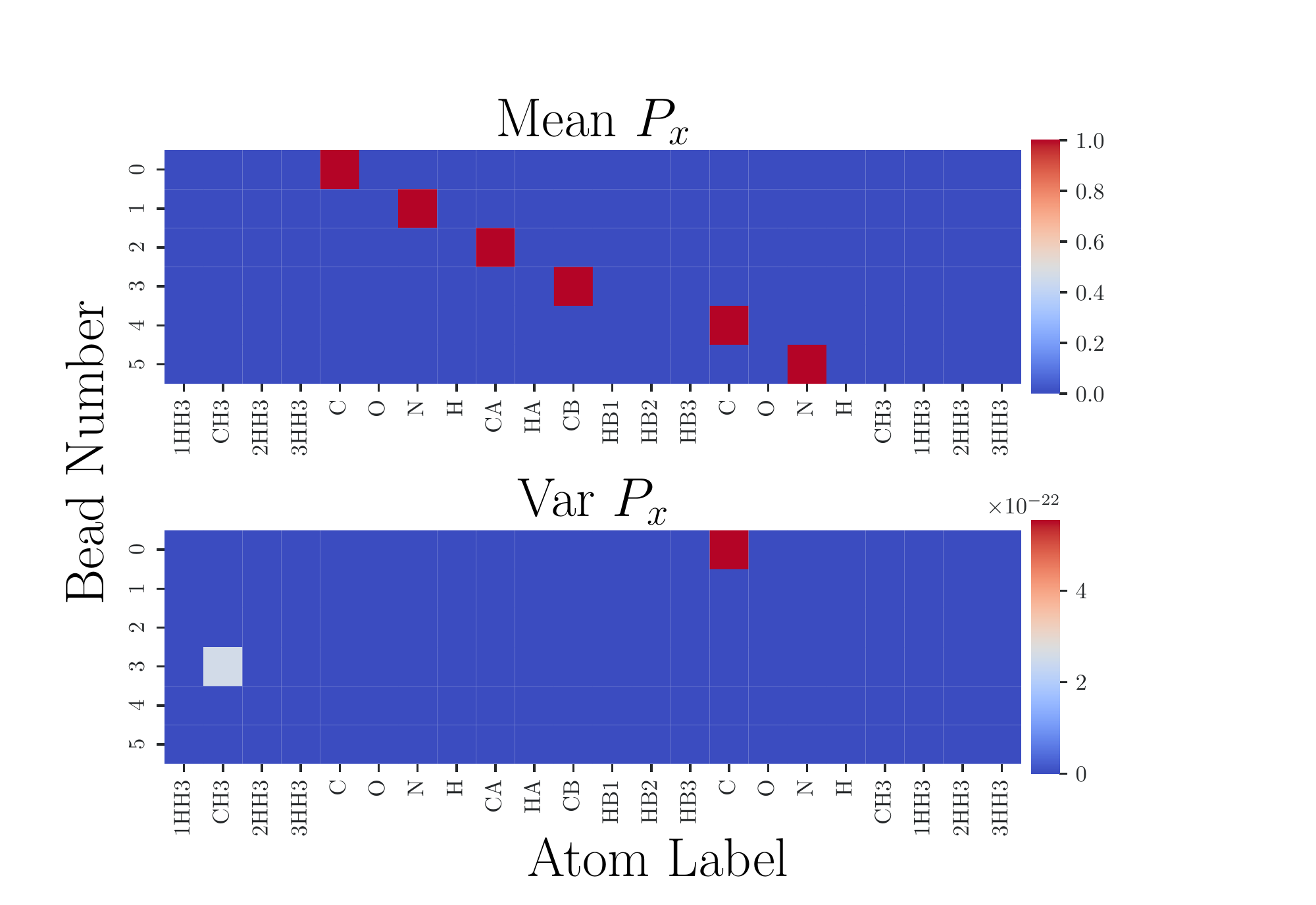}
    \caption{Alanine dipeptide. Projection Matrices $P_{\xb} = \Theta(\xb)$ computed for all fine-grained configurations $\xb$ in dataset. Mean $P_{\xb}$ (top) and Variance $P_{\xb}$ (bottom). Coarse-graining maps are essentially the same across all fine-grained configurations in dataset and heavily weigh backbone atoms. Bead atoms are reindexed for ease of visualization such that lower bead indices correspond to C-terminus side and higher bead indices correspond to N-terminus.}
    \label{fig:adp_proj}
\end{figure}

\begin{figure}
    \centering
    \includegraphics[width=1.25\linewidth]{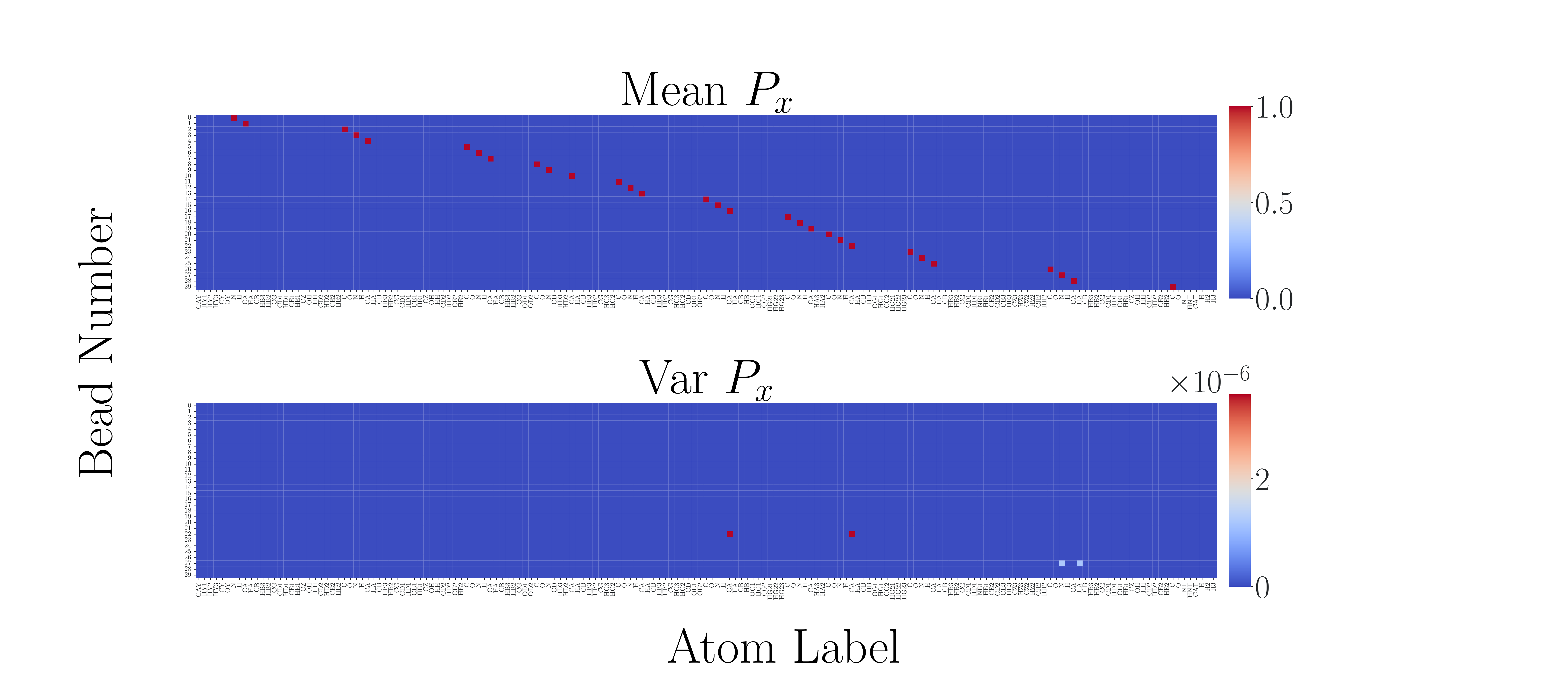}
    \caption{Chignolin. Projection Matrices $P_{\xb} = \Theta(\xb)$ computed for all fine-grained configuration $\xb$ in dataset. Mean $P_{\xb}$ (top) and Variance $P_{\xb}$ (bottom). Coarse-graining maps are essentially the same across all fine-grained configurations in dataset and heavily weigh backbone atoms. Bead atoms are reindexed for ease of visualization such that lower bead indices correspond to C-terminus side and higher bead indices correspond to N-terminus.}
    \label{fig:chig_proj}
\end{figure}

We plot the mean and variance of all the projection matrices computed for each configuration in our dataset for alanine dipeptide (Figure~\ref{fig:adp_proj}) and chignolin (Figure~\ref{fig:chig_proj}). For each of the systems we investigate here, the projection matrices learned are essentially the same across for all the configurations in dataset. Importantly, the learned coarse-graining map corresponds to a physically meaningful map, where backbone atoms $(\rm{C}, \rm{C}_{\alpha}, \rm{N})$ are given a high weight, with side chain atoms given a zero weight. The conformational dynamics of the two proteins we investigate are dominated by backbone behavior and the coarse-graining maps we use respect this behavior. For more complex biomolecular systems and/or for a coarse-graining aimed at a greater dimensionality reduction, the coarse-graining map will be less physically obvious. The approach we introduce here remains robust to these potential challenges.

\begin{table*}
\centering
\begin{tabular}{|p{5cm}||p{4cm}|p{4cm}|}
 \hline
 \multicolumn{3}{|c|}{Hyper-Parameters} \\
 \hline
 Hyperparameter & Chignolin & Alanine Dipeptide\\
 \hline
 $k$ ($\#$ of Beads) & 30 & 6 \\
 $\bar{n}$ ($\#$ of Target Atoms) & 40 & 13 \\
 $\alpha_{\rm bb}$ & 5.0 & 2.0 \\
 $\hat{U}$ Optimizer & Adam  & Adam\\
 $\hat{U}$ Learning Rate (LR)  & $3 \times 10^{-4}$    &$3 \times 10^{-4}$\\
 $\hat{U}$ Scheduler & ReduceLROnPlateau   & ReduceLROnPlateau\\
 $\hat{U}$ Patience & 5   & 5\\
 $\hat{U}$ Factor & 0.8   & 0.8\\
 $\hat{U}$ Minimum LR & $1 \times 10^{-6}$   & $1 \times 10^{-6}$\\
 $\hat{U}\  \#$ of $\rm{SchNet}$ layers & $2$   & $2$\\
 $\hat{U}$  width & $128$   & $128$\\
 $\hat{U}$  cutoff (\AA) & $15$   & $10$\\
 $\hat{U}\  \nb_{\rm gaussians}$  & $25$   & $25$\\
 $\Theta$ Optimizer & Adam  & Adam\\
 $\Theta$ Learning Rate (LR)  & $1 \times 10^{-4}$    &$1 \times 10^{-4}$\\
 $\Theta$ Scheduler  & ReduceLROnPlateau   & ReduceLROnPlateau \\
 $\Theta$ Patience & 5   & 5\\
 $\Theta$ Factor & 0.8   & 0.8\\
 $\Theta$ Minimum LR & $1 \times 10^{-6}$   & $1 \times 10^{-6}$\\
 $\Theta\  \#$ of $\rm{E(\emph{n})}$ layers  & $2$   & $2$\\
 $\Theta$ width  & $128$   & $2$\\
 $\Theta$ nonbonded-edge cutoff (\AA) & $5$   & $2$\\
 $\lambda$ & 5.0 & 0.2 \\
 $\lambda_{\rm mf}$ & 0.001 & 0.001 \\
 $\nb_{\rm freeze\_cg}$ & 2 & 5 \\
 $\nb_{\rm cg}$ & 15 & 10 \\
 $\nb_{\rm u}$ & 10 & 10 \\
 Batch Size & 8 & 8 \\
 Number of Datapoints & 50000 & 50000 \\
 \hline
\end{tabular}
\caption{Relevant hyperparameters for encoder-decoder ($\Theta-\Theta^{\rm dec}$) and $\hat{U}$ training}
\label{tab:cgu}
\end{table*}

\begin{table*}

\centering
\begin{tabular}{|p{4cm}||p{5cm}|p{3cm}|}
 \hline
 \multicolumn{3}{|c|}{Hyper-Parameters} \\
 \hline
 Hyperparameter &Chignolin&Alanine Dipeptide\\
 \hline
 RQ-NSF Layers & 2  & 2\\
 Width  & 256   & 128\\
 Optimizer & Adam  & Adam\\
 Learning Rate  & $3 \times 10^{-4}$    &$3 \times 10^{-4}$\\
 Number of Datapoints & 50000 & 50000 \\
 Batch Size & 128 & 64 \\
 \hline
\end{tabular}
\caption{Relevant hyperparameters for normalizing flow training}
\label{tab:nf}
\end{table*}

\section{Coarse-Grained Simulations}\label{app:cgsim}
Given a trained coarse-grained potential function $\hat{U}$, we can sample the coarse-grained space via standard Langevin dynamics. We carry out dynamics using an ``OVRVO'' integration scheme \cite{sivak_time_2014}. We use the $\texttt{OpenMM}$ \cite{eastman_openmm_2017} simulation platform to carry out all coarse-grained simulations. Finally, we use the $\texttt{TorchForce}$ plugin to interface our coarse-grained potential energy function $\hat{U}$ into the simulation.

Given a vector $\mb = [m^1, m^2, ... m^n]$ consisting of all the masses of the atoms in the fine-grained configuration, we define the mass of a bead to be $\mb \left<P_{\xb}\right>_{\mathcal{D}}^T$, where $\left<P_{\xb}\right>_{\mathcal{D}}$ is the average projection matrix across all configurations in our dataset $\mathcal{D}$. For the coarse-graining maps we learn, our projection matrices are essentially the same across all fine-grained configuration in $\mathcal{D}$, so this approach is a sensible one. For systems with more complex coarse-graining maps, a different strategy to account for masses will likely be necessary.

For alanine dipeptide, we ran 6 coarse-grained trajectories each for a total of 2000000 steps. We used a time step of $\rm{dt}=0.0001 \rm{ps}$ and friction coefficient $\gamma=100\rm{ps}^{-1}$. Three of the trajectories were started in the top right basin and three of the trajectories were started in the right most basin. We subsample $20000$ total coarse-grained configuration from these trajectories  leading to a total of 20000 coarse-grained configurations. We then carry out our back-mapping procedure and for each coarse-grained configuration we generate 200 points for a total of 4000000 fine-grained configurations. Finally, we carry out a reweighting step followed by a relaxation step detailed below. 

For chignolin, we initially ran 24 trajectories each for a total of 2000000 steps. For these trajectories, we used a time step of $\rm{dt}=0.001 \rm{ps}$ and friction coefficient $\gamma=100\rm{ps}^{-1}$. We started 8 of these trajectories in the first basin (unfolded state), 8 of these trajectories in the second basin (folded state), and 8 of these trajectories in the third basin (misfolded state). When carrying out the back-mapping process, we observed that trajectories that started in the misfolded state quickly transitioned to one of the other 2 basins, resulting in limited sampling of the misfolded state. To alleviate this, we ran 3 more trajectories starting in this basin with a $\rm{dt}=0.00001\rm{ps}$ and friction coefficient $\gamma=100\rm{ps}^{-1}$, also of length 2000000 steps. From these 27 trajectories, we back-map 54000 coarse-grained configurations and generate 750 fine-grained configuration per coarse-grained configuration. Finally, we carry out a relaxation step followed by a reweighting step detailed below.

\section{Reweighting Procedure}
\label{app:reweighting}
We sample from the Boltzmann distribution via a sampling scheme that combines coarse-grained dynamics with a back-mapping procedure that utilizes a normalizing flow. There is no theoretical guarantee that this sampling scheme allows us to sample from the true Boltzmann distribution; instead, we must carry out a final reweighting to ensure that samples are appropriately weighted. Below, we highlight two different reweighting procedures, one that explicitly uses $p(\zb)$ and $p(\xb|\zb)$ the other that implicitly accounts for it. To limit the computational burden, all energies are computed via an implicit model using $\texttt{OpenMM}$. In addition, we carry out a short relaxation procedure of overdamped Brownian dynamics---again using an implicit model---in order to relax any minor structural deformities and ensure configurations are reflective of ambient temperature.
\subsection{Alanine dipeptide}
Given a fine-grained configuration $x_j$ back-mapped from a coarse-grained configuration $\zb_i$ (i.e. $\xb_j \sim p(\xb | \zb_i)$), we weigh each configuration according to $w_{ij}/\sum_{ij}(w_{ij})$, where $w_{ij} = \frac{\exp(-\beta U_{\rm implicit}(\xb_j))}{{\exp(-\beta \hat{U}(\zb_i))*p(\xb_j|\zb_i)}}.$ The sum is taken over all fine-grained configurations $\xb_j$ back-mapped from all coarse-grained configurations $\zb_i$. Finally, we carry out a single relaxation step of overdamped Brownian dynamics with a time-step of $0.00002\ \rm{ps}$ and friction coefficient $\gamma=10\ \rm{ps}^{-1}$. The time-step we use for this relaxation is 200$\times$ lower than the time-step used to carry out the original MD simulation.

\subsection{Chignolin}
For chignolin, the fine-grained structures we back-map are generally high-energy
configurations; however, this was generally a result of minor structural deformities as opposed to major flaws in the reconstruction procedure. Because of this high-energy, we cannot carry out the same reweighting procedure we used with alanine dipeptide. However, we do have access to the distribution of potential energies---recomputed using an implicit solvent model---of the configurations in our MD-dataset, which we reweight with respect to. First, we carry out 300 steps of overdamped Brownian dynamics with a time-step of $0.00002\ \rm{ps}$ and friction coefficient $\gamma=10\ \rm{ps}^{-1}$ in order to relax the structure to a reasonable energy. 

Then, we reweight the distribution of potential energies from our relaxed configurations with respect to distribution of potential energies from our MD dataset. Given a histogram binning function $h: \RR \to [0, 1]$ that computes the probability of the bin a potential energy belongs to, we assign to each fine-grained configuration $\xb_j$ back-mapped from a coarse-grained configuration a weight $w_{ij}/\sum_{ij}(w_{ij})$, where $w_{ij} = \frac{h^{\rm MD}(U_{\rm implicit}(\xb_j)}{h^{\rm gen}(U_{\rm implicit}(\xb_j)}$.

\section{Datasets}
We carried out a simulation of alanine dipeptide in explicit solvent at a temperature of 300K with the AMBER ff99SB force field and the TIP3P water model. We used a time-step of $0.004 \rm{ps}$ and a friction coefficient of  $\gamma=0.1\rm{ps}^{-1}$.
The total simulation time was over 0.5 $\mu$s. Finally, we saved the configurations every 1 ps for a dataset of over 500000 points. From this dataset, we subsampled 50000 data points consisting of positions and forces, which were then used for training.

We used the dataset of chignolin trajectories from \cite{husic_coarse_2020}. See supplement in \cite{husic_coarse_2020} for simulation details. From this dataset, we similarly subsampled 50000 data points consisting of positions and forces to carry out our training

\section{Weak thermodynamic consistency}

In this appendix, we state and prove the proposition stated in the main text. 

\begin{definition}
An invertible coarse-graining is a tuple $(\Theta, \hat U, T)$
consisting of coarse-graining map $\Theta:\RR^{3n} \to \RR^{3k}$, a coarse-grained potential $\hat U: \RR^{3k}\to \RR$, and a normalizing flow $T:\RR^{3n}\to\RR^{3n}$. 
\end{definition}

\begin{definition}
A projective coarse-graining is a pair $(\Theta, \hat U)$ consisting of a coarse-graining map $\Theta: \RR^{3n} \to \RR^{3k}$ and a coarse-grained potential $\hat U: \RR^{3k} \to \RR$.
\end{definition}

\begin{definition} \label{def:thcs}
We call a projective coarse-graining $(\Theta, \hat U)$ thermodynamically consistent if
\begin{equation}
    \hat U(\zb) = -\beta^{-1} \log \int \delta (\Theta(\xb) - \zb) e^{-\beta U(\xb)} d\xb.
\end{equation}
\end{definition}

\begin{proposition}
Let $(\Theta, \hat U, T)$ be an invertible coarse-graining. 
Let $\mathcal{F}_*$ denote the set of functions of continuous, bounded functions, 
$$ \mathcal{F}_* := \{ f \in \mathcal{C}(\RR^{3n},\RR) \big| \avg{f}_{\xb} = \avg{f(\xb) \delta(\Theta(\xb) - \zb)}_{\xb, \zb} \},$$
where $\avg{\cdot}_{\xb}$ denotes an ensemble average with respect to the fine-grained Boltzmann distribution $\rho(\xb) d\xb = e^{-\beta U(\xb)} d\xb$ and $\avg{\cdot}_{\xb,\zb}$ is also integrated over $\zb$. 
If $(\Theta, \hat U, T)$ is $\mathcal{F}_*$ thermodynamically consistent, then the projective coarse-graining $(\Theta, \hat U)$ is thermodynamically consistent in the sense of Ref.~\cite{noid_multiscale_2008}.
\end{proposition}

The proof of this proposition is straightforward; we simply use $\mathcal{F}_*$ weak thermodynamic consistency in the fine-grained space and project. The assumption we make is that the degrees of freedom orthogonal to $\Theta(x)$ do not contribute to the average for functions in $\mathcal{F}_*.$ 
First, let $f\in \mathcal{F}_*.$ By assumption, 
\begin{equation}
   \avg{f} = \int f(\xb) e^{-\beta U(\xb)} d\xb = \int f(\xb) p_{\rm gen}(\xb | \zb) e^{-\beta \hat U(\zb)} d\zb d\xb
\end{equation}
We also have assumed that for all $f\in \mathcal{F}_*$,
\begin{equation}
 \int f(\xb) e^{-\beta U(\xb)} d\xb = 
 \int f(\xb) e^{-\beta U(\xb)} \delta (\Theta(\xb) - \zb) d\xb d\zb
\end{equation}
so, in particular, 
\begin{equation}
 \int f(\xb) e^{-\beta U(\xb)} \delta (\Theta(\xb) - \zb) d\xb d\zb = \int f(\xb) p_{\rm gen}(\xb | \zb) e^{-\beta \hat U(\zb)} d\zb d\xb
\end{equation}
which implies that for any $\hat f:\RR^{3k}\to \RR$
\begin{equation}
\begin{aligned}
 \int \hat f(\Theta(\xb)) e^{-\beta U(\xb)} \delta (\Theta(\xb) - \zb) d\xb d\zb &=  \int \hat f(\Theta(\xb)) p_{\rm gen}(\xb | \zb) e^{-\beta \hat U(\zb)} d\zb d\xb \\
 &= \int \hat f(\zb) e^{-\beta \hat U(\zb)} d\zb \\
 \end{aligned}
\end{equation}
which follows from the normalization of $p_{\rm gen}.$
The equality obtained here is the weak form of~\ref{def:thcs}. 
To obtain pointwise convergence, we construct a sequence of Gaussian distributions centered at each $\zb$ that approach a $\delta$-function.

\end{document}